\documentclass[10pt,twocolumn]{article}
\usepackage{booktabs} 
\usepackage[ruled]{algorithm2e} 

\SetAlFnt{\small}
\SetAlCapFnt{\small}
\SetAlCapNameFnt{\small}
\SetAlCapHSkip{0pt}

\usepackage{float} 
\usepackage{multirow} 
\usepackage{amsmath,amssymb,bm,mathrsfs}
\usepackage[utf8x]{inputenc}
\usepackage{caption}
\usepackage{tabularx}

\pdfoutput=1                     

\usepackage[margin=0.75in,bottom=1in]{geometry}            
\usepackage{setspace} 
\usepackage{newtxtext,newtxmath}             
\usepackage[T1]{fontenc}
\usepackage[utf8x]{inputenc}

\usepackage{graphicx}
\usepackage{float}
\usepackage{booktabs,tabularx,multirow}
\usepackage{caption,subcaption}
\usepackage{algpseudocode}
\usepackage{authblk}
\usepackage{amsmath,amssymb,bm,mathrsfs}

\usepackage[square,numbers]{natbib}
\usepackage[colorlinks=true,
            linkcolor=blue,
            citecolor=blue,
            urlcolor=blue]{hyperref}

\newcommand{\shortauthors}[1]{}              
\newcommand{\keywords}[1]{\paragraph*{Keywords} #1}

\author{Narayan Kandel}
\author{Daljit Singh J.S. Dhillon}

\affil{School of Computing, Clemson University, Clemson, SC, USA}

\date{}

\begin{document}
\title%
     {GratNet: A Photorealistic Neural Shader for Diffractive Surfaces}     
\newcommand{\FLIP}{\reflectbox{F}LIP }
\maketitle
\begin{figure*}
\centering
    \begin{tabularx}{\linewidth}{@{}c@{}c@{}c@{}c@{}c@{}}
    \includegraphics[width=0.199\textwidth]{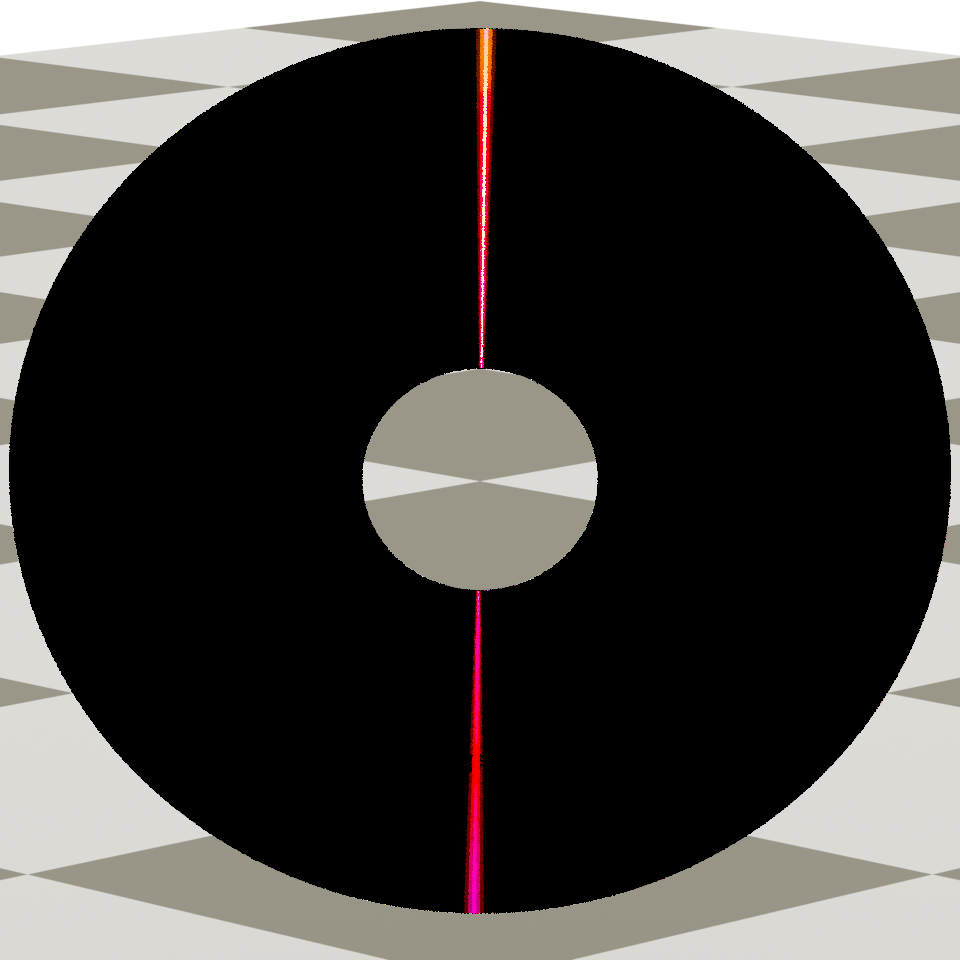} &
    \includegraphics[width=0.199\textwidth]{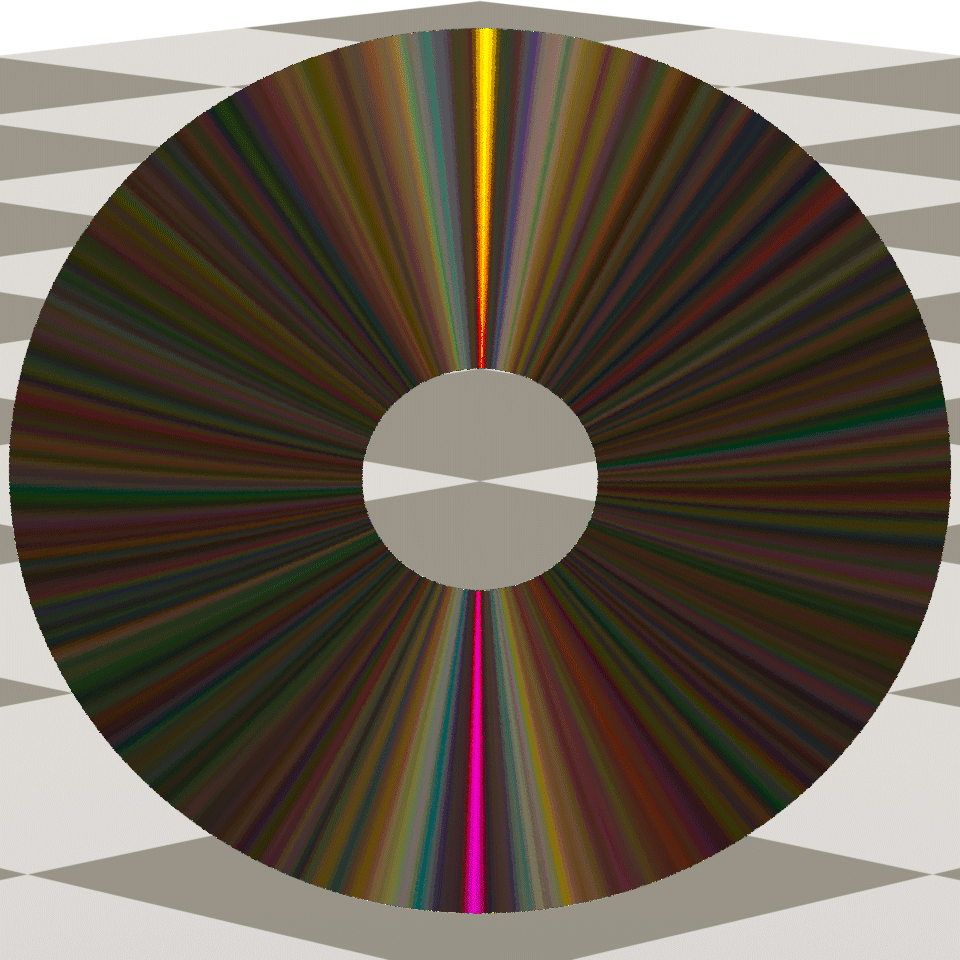} &
    \includegraphics[width=0.199\textwidth]{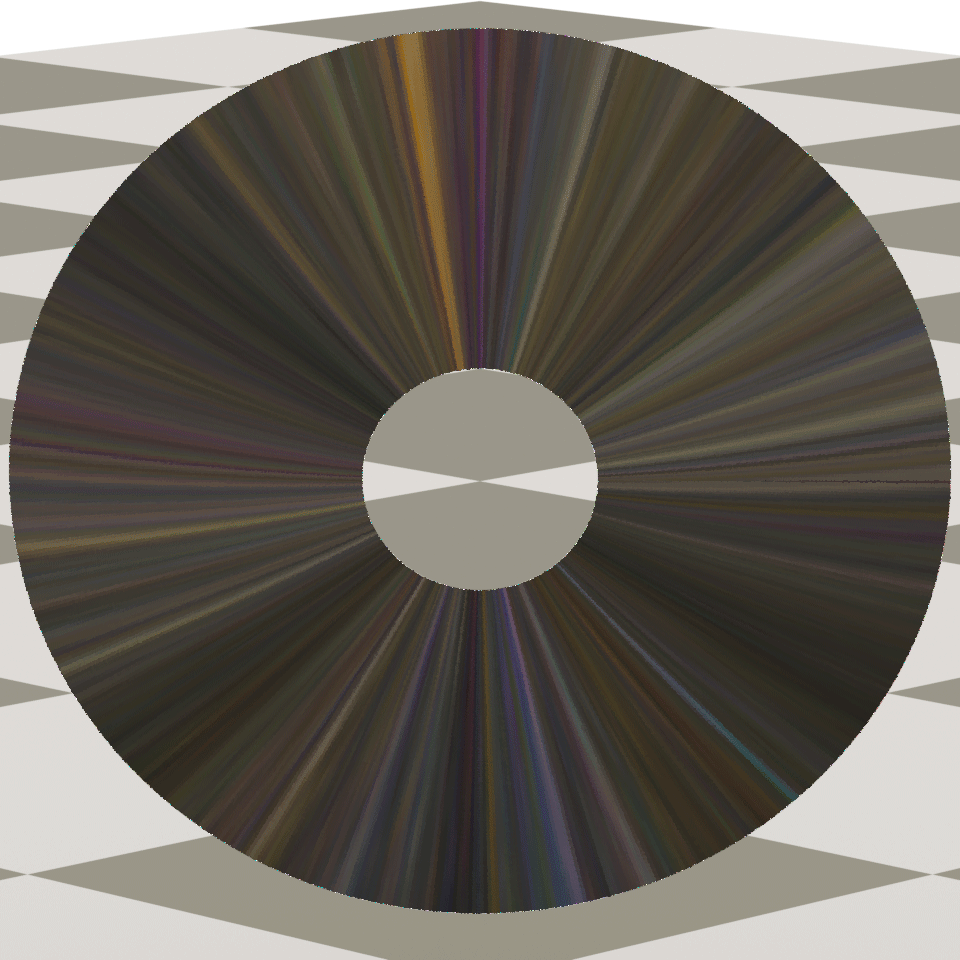}&
    \includegraphics[width=0.199\textwidth]{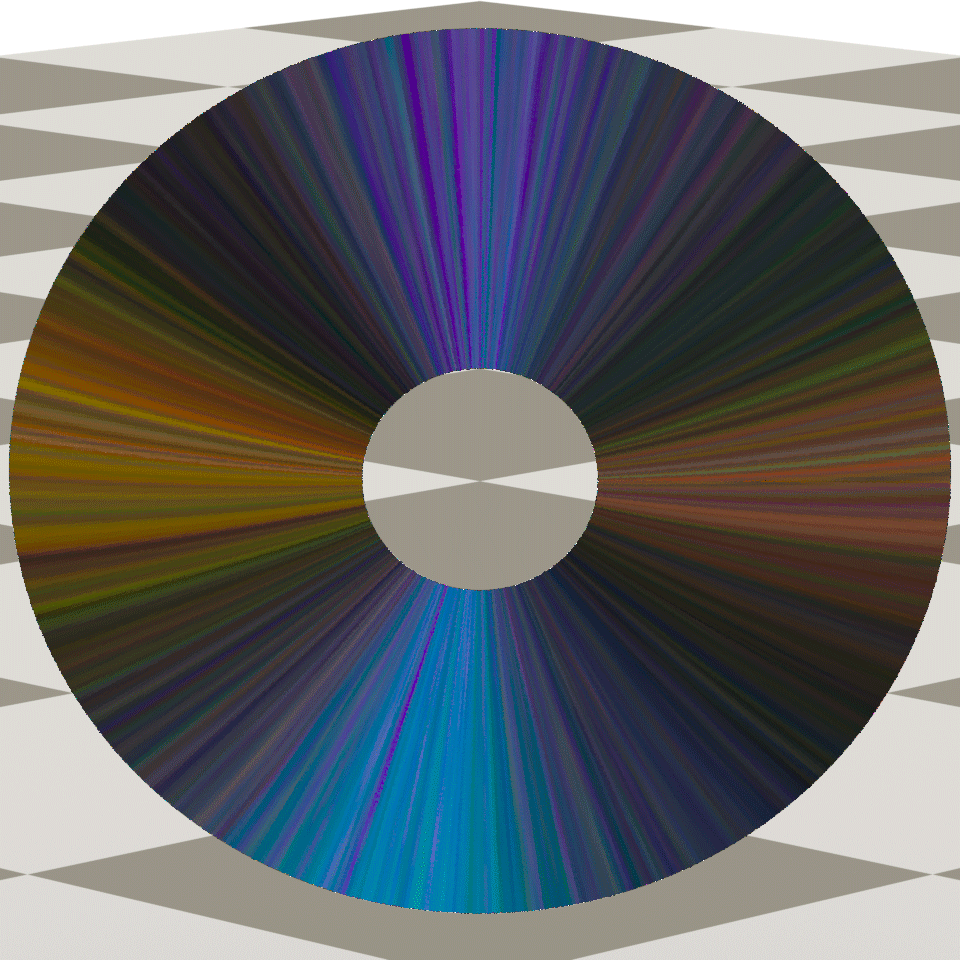}&
    \includegraphics[width=0.199\textwidth]{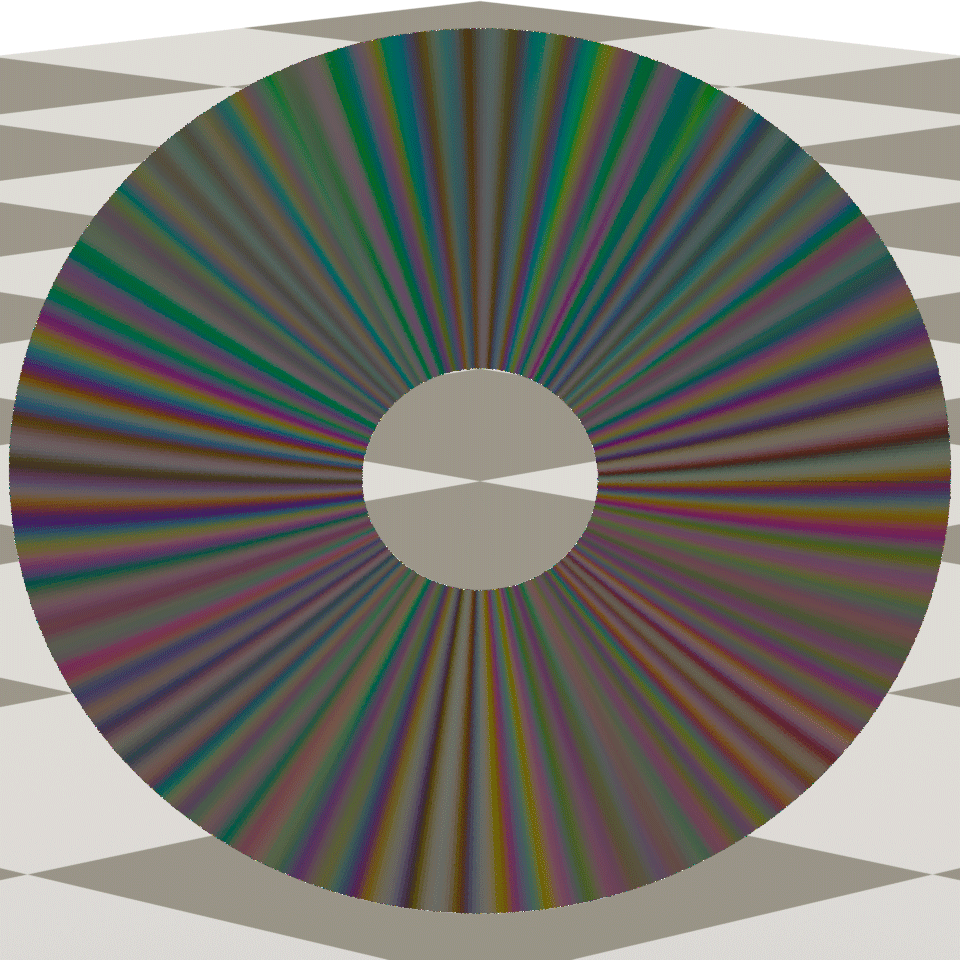}\\
   (a) Holographic Grooves & (b) CD Pits \& Lands &
    (c) Metal Scratches & (d) Snake Skin~ & (e) Spherical Pits 
    \end{tabularx}
    \caption{Structural colors arising from diffraction by various surface nanostructures. We emulate a real-world sized compact disk with concentric tracks. Nanostructures oriented along those tracks are depicted in an order of increasing complexity from the left to right. (a) Sinusoidal holographic grooves with $600$ lines-per-mm, (b) CD-ROM with random pits and lands on tracks that are $1.5\mu m$ apart, (c) metallic scratches on the tip of a ball-pen~\cite{nanoSurf}, (d) natural V-shaped grooves on a sunbeam snakeskin~\cite{dhillon2014}, and (e) highly complex nanostructures with semi-spherical pits of mild random size and placement variations~\cite{yu2023} are depicted here. The rainbow-\textit{ish} radial bars of near equal thickness and regularity in (e) result from a reflectance that is \textit{difficult} to emulate~\cite{yu2023}. The snakeskin and metallic scratches are real-world atomic-force microscopic scans while the rest of the nanostructures are synthetic. We generate reference \textit{ground-truth} using Kirchhoff's formula to emulate scalar-wave Fourier optics for far-field Fraunhofer diffraction. Our neural shader learns and reconstructs diffractive reflectances of all levels of complexity with high degree of accuracy as presented in Section~\ref{sec:Results}. Our method is realized as a deferred shader for \textit{Mitusba3} renderer~\cite{mitsuba3}. To aid visual comparisons of structural coloration across the examples, we use the same base dielectric material, exclude Fresnel effects that become uniform across the cases and render scratches at $10\times$ the exposure in comparison to the others. 
    }
    \label{fig:teaser}
  \end{figure*}

\begin{abstract}
\small
\it
Structural coloration is commonly modeled using wave optics for reliable and photorealistic rendering of natural, quasi-periodic and complex nanostructures. Such models often rely on dense, preliminary or preprocessed data to accurately capture the nuanced variations in diffractive surface reflectances. This heavy data dependency warrants implicit neural representation which has not been addressed comprehensively in the current literature. In this paper, we present a multi-layer perceptron (MLP) based method for data-driven rendering of diffractive surfaces with high accuracy and efficiency. We primarily approach this problem from a data compression perspective to devise a nuanced training and modeling method which is attuned to the domain and range characteristics of diffractive reflectance datasets. Importantly, our approach avoids overfitting and has robust resampling behavior. Using Peak-Signal-to-Noise (PSNR), Structural Similarity Index Measure (SSIM) and a flipping difference evaluator (\FLIP) as evaluation metrics, we demonstrate the high-quality reconstruction of the ground-truth. In comparison to a recent state-of-the-art offline, wave-optical, forward modeling approach, our method reproduces subjectively similar results with significant performance gains. We reduce the memory footprint of the raw datasets by two orders of magnitude in general. Lastly, we depict the working of our method with actual surface renderings.

\end{abstract}

\section{INTRODUCTION}
\label{sec:Intro}
Material appearance modeling is critical to photorealistic rendering of VFX effects. Many diffractive surfaces are commonly present in virtual scenes and warrant efficient modeling and shading.
Surface nanostructures diffract light waves of different wavelengths differently to produce \emph{structural colors}. These specular colors impose a challenging problem with \textit{iridescent} reflectances (and transmittances) where a fixed surface point changes color depending on the incident light or the view direction. This problem is further exacerbated by natural variations in surface nanostructures. Fine examples of such diffractive coloration include flower petals, butterfly wings, rocks, sea-shells as well as snake bodies. Also, man-made objects like compact disks, holographic papers, scratched metallic sheets, and etched windowpanes exhibit nuanced diffractive iridescence due to their design, fabrication imperfections or wear and tear.

Periodic, well-defined diffractive gratings can be modeled analytically. However, natural variations or complex fabrications discussed above introduce a \textit{non-analytic} function in representing the surface profile. This makes it impossible to mathematically deduce a closed-form formulation for directly computing their reflectances. State-of-the-art methods based on wave optics exploit ways to discretize such non-analytic surface profiles to enable numerical approximations~\cite{yu2023, yanRavi2018, dhillon2014}. Such forward modeling approaches often prove to be memory and/or computationally intensive. \citet{yu2023} has emphasized that for practical usage of their highly accurate full-wave simulator, significant performance improvements are warranted. Also,
the accuracy of such forward modelers heavily depends on the richness of the data representing the surface profiles. 
In crux, a data-driven approach is warranted to assimilate the statistical measures arising from the underlying stochastic light-matter interactions~\cite{stam1999}.      

Neural network modeling has the potential to remedy the discussed limitations of computational load and memory footprint, simultaneously. 
This is mainly due to their capacity and efficiency in effectively learning the nuanced variations in training datasets. Yet, using neural networks for modeling diffractive reflectances has largely been limited to sheen effects in specular highlights, glints~\cite{zhu2022recent} and irregular surface scratches. 
\citet{kuznetsov2019} rely on convolutional neural networks (CNNs) to model spatially varying glinty scratches. 
CNNs for diffraction shading are computationally demanding and have a large memory footprint. 
In a broader context, low footprint, multi-layer perception (MLP) have efficiently modeled isotropic as well as non-isotropic reflectances (BRDFs)~\cite{timWeyrich2021}, bidirectional texture functions (BTFs)~\cite{kuznetsov2021NeroMip} and even real-time neural appearances under global illumination~\cite{zeltner2024}. 
Yet, to the best of authors' knowledge, no existing implicit neural representation based method can reproduce the nuances of rich diffractive reflectances as modeled using wave-optics~\cite{yu2023,dhillon2014}. Key challenges to such emulations arise due to the scale, volume and diversity of the nano-structural variations under the surface footprint of each display pixel.  
In this paper, we present a novel multilayer perceptron based method to effectively model nuanced diffraction from natural quai-periodic nanostructures as well as complex man-made nano-architectures. 
Inspired by the recent progress in image compression domain~\cite{catania2023nif,strumpler2022implicit,tancik2020fourier} and the notion of data-compression in implicit representation of appearance models~\cite{timWeyrich2021}, we devise our method to address the unique domain and range space representational challenges of diffractive reflectance data-sets. 
The key ideas and contributions in this paper include:
\begin{itemize} 
\setlength{\partopsep}{0pt}
\setlength{\topsep}{0pt}
\setlength{\parsep}{0pt}
\setlength{\itemsep}{5pt}
\item a data transformation and re-organizational scheme specialized to improve the learnability of diffractive reflectances,
\item an approach to construct MLP layers with funneling under golden ratio to actuate annihilated learning, and
\item A method that delivers high accuracy even for the most complex cases~\cite{yu2023} with major performance gains and significant memory footprint reductions in general.

\end{itemize}  
 
\vspace{-5mm}     
\section{RELATED WORK}
\label{sec:Related}

\paragraph*{Forward Modeling} In physics, structural coloration in off-surface reflectances is explained with the electromagnetic wave theory. Particular roles of wave properties such as wavelength, polarization, and \emph{coherence} are accurately expressed through Fourier optics in modeling (far-field) Fraunhofer diffraction. Some of the earliest efforts in bringing wave optics to computer graphics are well documented by Stam in his seminal work  based on Kirchhoff's theory~\cite{stam1999}. He applies it to scalar light waves for modeling diffraction in simpler, 'statistically' parameterizable cases like rainbow colors on compact disks and metallic specular highlights. Yet, it offers a solid physics-based foundation for emulating surface diffraction using analytical surface profiles as well as statistical estimates on them. Several follow-up research expanded the horizon of forward modeling to global illumination \cite{cuypers2012}, accurate simulation of natural gratings on animal skins~\cite{dhillon2014}, metallic scratches~\cite{werner2017}, speckles~\cite{yanRavi2018} and pearlescent surfaces~\cite{pearl2020}. Multilayer diffraction~\cite{bragg2024, dhillon2021} and multiscale interplays can also be forward modeled accurately. More recently, Yu et al.~\cite{yu2023} developed a Boundary-Element based full wave simulator that is more accurate than Kirchhoff-based~\cite{stam1999,dhillon2014,yanRavi2018} or the alternative generalized Harvey-Shack theory based method~\cite{krywonos2006predicting} in addressing unusual surface geometries. In the context of global illumination, Steinberg and colleagues have developed a series of methods that bundle wave properties and resolve wave-matter interactions through some form of radiance generalization and respective mathematical simplifications~\cite{steinberg2021generic,Steinberg_practical_plt_2022, steinberg2024}. For the nano-architectures addressed in this paper, only few state of the art forward methods can generate accurate results~\cite{yu2023, steinberg2021generic, dhillon2014, krywonos2006predicting} and except one~\cite{dhillon2014}, rest of them are not devised to be computationally inexpensive. Furthermore, all of them impose memory requirements that are particularly challenging for interactive pipelines. In this paper, we reply on such accurate methods to provide the ground-truth datasets that can then be efficiently and effectively modeled through implicit neural representations.      

\paragraph*{Inverse Methods} Acquiring high quality, large volume data for inverse modeling structural coloration is generally requires customized and expensive setups. In the last decade, few effective, imaging-based methods relying on consumer-grade cameras and DIY supplies have shown specific promises. Toisoul et al.~\cite{toisoul2017practical} 
develop a simple, image-based method to recreate surface diffraction colors using aperture induced \emph{bokeh}. Their method works for printed holographic materials with homogeneous distribution of highly regular gratings. Toisoul et al.~\cite{toisoulDhillon2018} explore the use of light polarization properties to extract two defining parameters for similar printed gratings with spatial variations, namely orientation and periodicity. Such methods are limited and cannot extract parametric representation of natural or complex surface nanostructures. Basic correlational statistics extractable from images are inadequate to represent their nuanced variations that are required for photo-realistic rendering. Such details are adequately obtained either by construction~\cite{yu2023,yanRavi2018,werner2017} or by sub-microscopic scanning~\cite{dhillon2014}. These methods establish the centrality of the datasets and their rigorous analysis under a numerical method, no matter how abstract,  to accurately recreate their appearance effects. 

\paragraph*{Implicit Neural Representations} One of the earliest implicit neural representations (INR) of diffractive reflectances used convolutional neural networks.  \citet{yanRavi2018} model spatially varying, near-random, diffractive coloration in scratches and glints. They employ CNNs to jointly model the coloration as well as microscopic variations. 
Designing neural networks for patterned structural coloration is a weakly explored area and thus their neural representation is an open problem. 
Away from structural coloration problems, simple multilayer perception based methods are used effectively in modeling a large variety of appearances as discussed in the introduction~\cite{timWeyrich2021,zhu2022recent,strumpler2022implicit}. These methods are not designed for address the range, scale and versatile domain distribution characteristics of the diffractive reflectance datasets. We empirically study these key characteristics. Resulting insights has helped us in devising effective data transformation and reorganization schemes to improve model learnability. In structuring and fine-tuning our MLP models, we look into generic data compression schemes dealing with image data.

\paragraph*{Data Compression}  MLP based Implicit Neural Representations (INRs) that use \textit{positional encoding} where the pixel location vector $(x,y)$ to query the network for the pixel color, can be very effective in compressing two-dimensional images~\cite{strumpler2022implicit}. ~\citet{timWeyrich2021} propose taking a data-compressive approach for implicit modeling of the reflectance data when it is available in a large enough volume. They devise a two-hidden-layer MLP network with inputs comprising of the Rusinkiewicz reparameterization~\cite{rusinkiewicz1998} of the light and the view directions. For the four-dimensional reflectance functions, this is akin to positional encoding, except that they use six instead of four input coordinates. This input feature  expansion aids the learning process where the input redundancy propagates implicit correlational constraints through the network. \citet{tancik2020fourier} apply frequency feature expansions to the positional codes to improve the learning of the high frequency features in the data. ~\citet{strumpler2022implicit} provide an elaborate survey of the state of the art in 2D image compression using INRs. More recently, \citet{catania2023nif} show that the use of sinusoidal activation functions~\cite{sitzmann2020implicit} and an auxiliary network that modulates the frequencies fed to these activators can achieve rate distortions that outperform traditional image compression encoders like JPEG and WebP at lower bitrates. However, they achieve these improvements by overfitting the data as their problem does not involve resampling. In our work, we aim at a simpler network structure for high performance. 
Also, our implicit representation must interpolate well in regions with no training data.
\section{PROPOSED METHOD}
\label{sec:Approach}

To devise an effective neural representational method for nuanced diffractive reflectances strategically, we: 
\begin{itemize}
    \item study the unique characteristics of such reflectance datasets
    \item devise appropriate range and domain space transformations
    \item experiment with the MLP neural network structure, size, activation, and the positional encoding schemes for the input, 
    \item validate with the larger set of real-world, synthetic as well as reference nano-structural examples, and  
    \item iteratively fine-tune the transformation, training, and representation modules for improving accuracy and performance of our method.
\end{itemize}

\noindent With the above strategy, we devise a novel method that includes two major steps. As the first step, for a given nanostructure profile, we generate the rich yet compact volume of its diffractive reflectance dataset based on our proposed range and domain space transformations. As the next step, we configure the network size and train it with a suitable sampling approach, as informed by our empirical studies. 
We present these two steps in detail but first the ground truth generation scheme is explained in the following.

\subsection{Generating the Ground Truth Training Datasets}
\label{sec:groundtruth}
With the focus on accuracy, we rely on a forward modeling technique to generate the ground-truth. We chose to implement a Fourier optics based shader that emulates Kirchhoff integral under dense sampling ($5nm$) of the visible spectral band. Our implementation is based on Stam's formulation~\cite{stam1999} as adapted by \citet{dhillon2014} for nanostructure profiles discretized into height-fields. \begin{figure*}[ht!]
\small
    \centering
    \begin{tabular}{c@{}c@{}c@{}c@{}c@{}c@{}c}
    \includegraphics[width=0.14\linewidth]{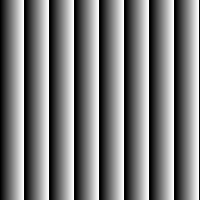} &
    \includegraphics[width=0.14\linewidth]{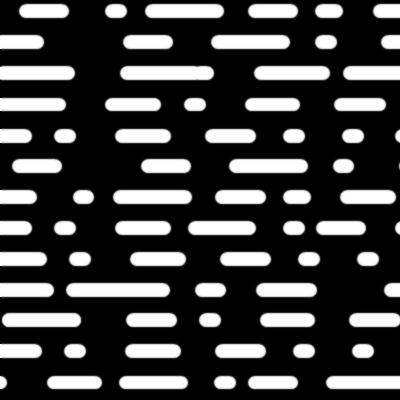} &
    \includegraphics[width=0.14\linewidth]{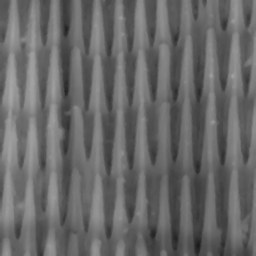} &
    \includegraphics[width=0.14\linewidth]{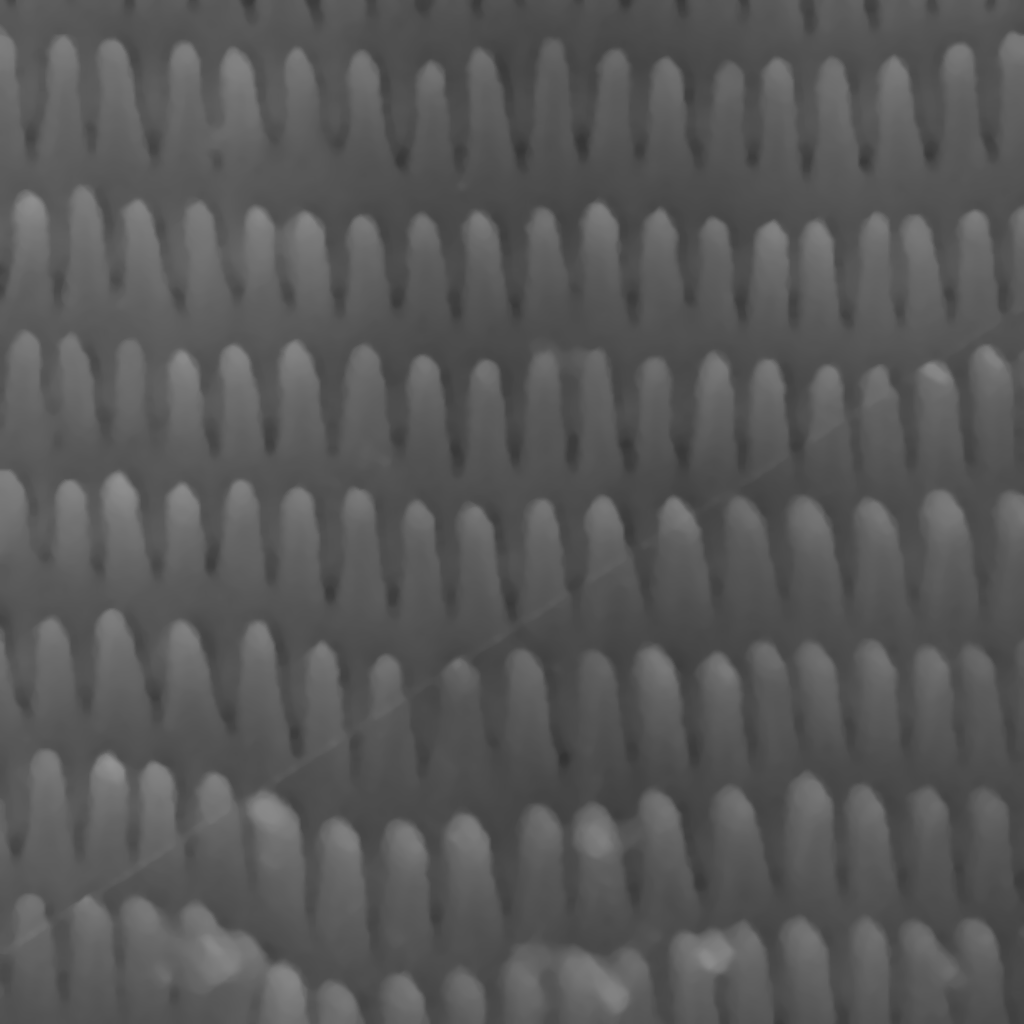} &
    \includegraphics[width=0.14\linewidth]{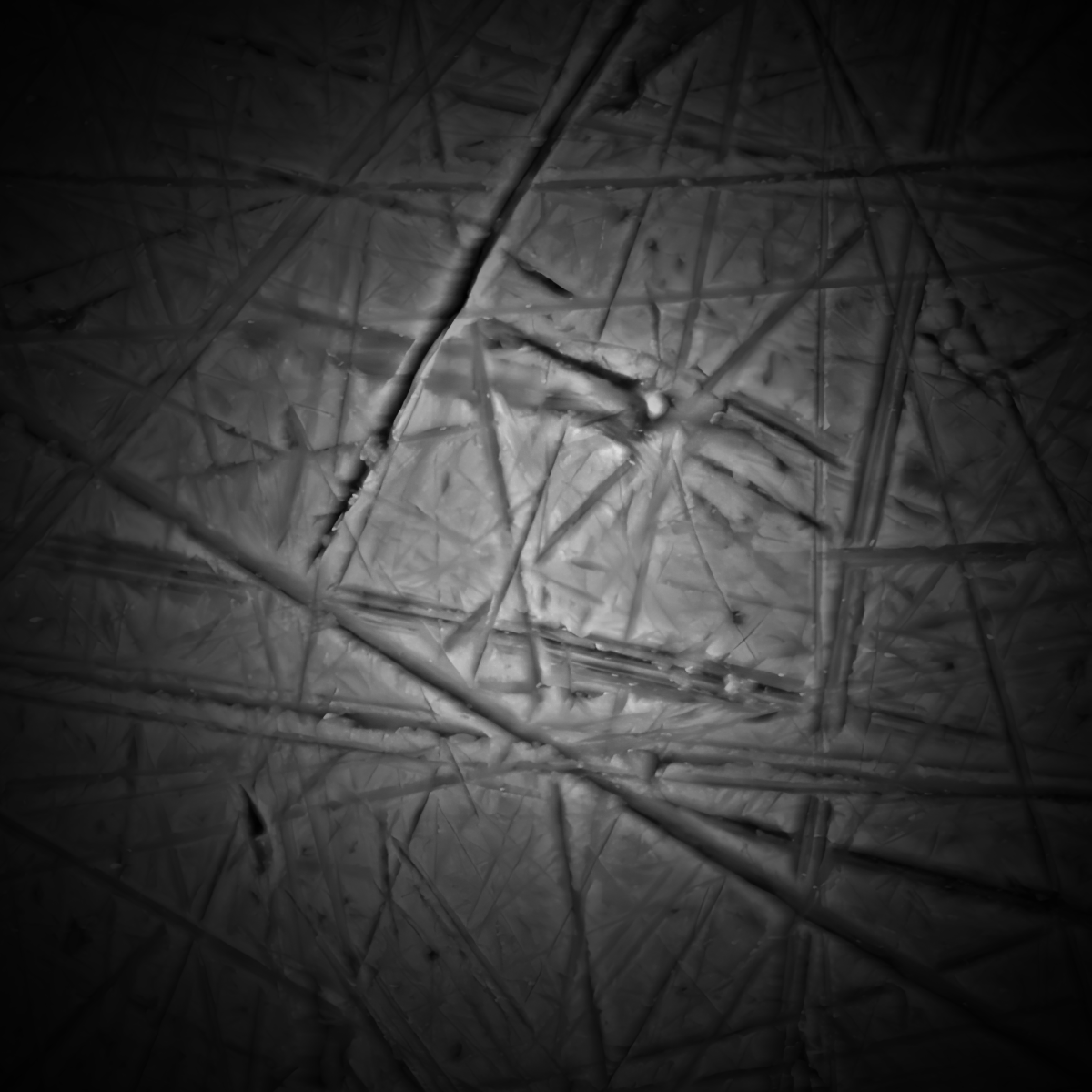} &
    \includegraphics[width=0.14\linewidth]{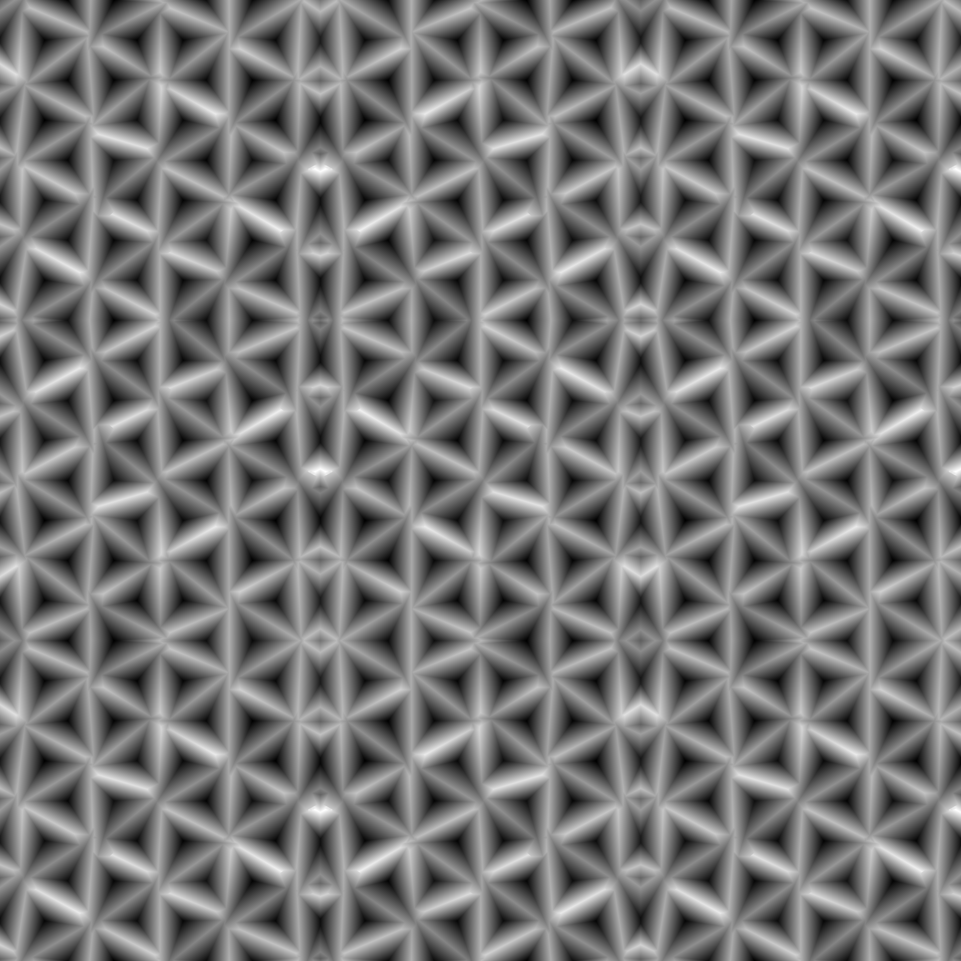} &
    \includegraphics[width=0.14\linewidth]{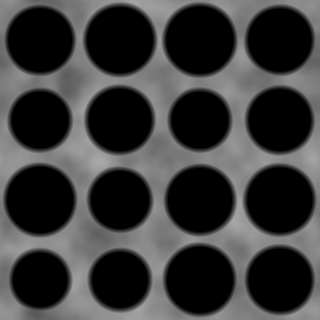} \\
    $20\mu m \times 20 \mu m$ &
    $20\mu m \times 20 \mu m$ &
    $12.5\mu m \times 12.5 \mu m$ &
    $10\mu m \times 10 \mu m$ &
    $80\mu m \times 80 \mu m$ &
    $72\mu m \times 72 \mu m$ &
    $24\mu m \times 24 \mu m$ \\
    Blazed Grating & Synthetic CD & Cornsnake & Sunbeam Snake & Ballpen Tip Scratch & Corner Cubes & Spherical Pits\\
    \end{tabular}

    \caption{Height-fields for Datasets. First two are synthetically generated by us. The snake scans were obtained by the courtesy of Milinkovitch, Teyssier, et al.~\cite{dhillon2014}. The ball-pen Tip Scans is by the courtesy  of NanoSurf Corporation~\cite{nanoSurf}. All the scans were generated through Atomic Force Microscopy (AFM) of real-world nanostructures. The last two are complex designs with nuanced variations and obtained by the courtesy of ~\citet{yu2023}. Actual patch sizes as used for forward modeling are bigger than or equal to the coherence window. Unless otherwise specified, we impose a coherence window of $4\sigma_s = 65 \mu m$. We use a coherence window of $4\sigma_s = 40 \mu m$ for Synthetic CD and Corner Cubes. $4\sigma_s = 15.556 \mu m$ for the Spherical Pits.    } 
    \label{fig:examples}
\end{figure*}
\vspace{5mm}

We represent nanostructures with a discrete 2D height-field function expressing nano-elevations, $h(x,y)$, along the $z$-axis. This is commonly done by existing forward modelers~\cite{yu2023,yanRavi2018,werner2017}. Some examples are shown in Figure~\ref{fig:examples}. Now, the Fourier optical form for its spectral bidirectional reflectance distribution function (BRDF) is given as:
\begin{align}
    \mathit{f}(\lambda, \omega_i,\omega_o) = A(\omega_i,\omega_o)\left| \mathcal{F}\{p(x,y)\}\left(u / \lambda,v / \lambda\right)\right|^2 ,
    \label{eqn:DTFT}
\end{align}
where, $\lambda$ is the light wavelength, $\omega_i$ is the normalized incident light direction and $\omega_o$ is the normalized view direction. The un-normalized half vector $\omega_h = \omega_i + \omega_o$ is related to Stam's key vector $(u,v,w) = -\omega_h$~\cite{stam1999}. 
$\mathcal{F}$ represents Discrete-Time Fourier Transform (DTFT) of the phasor function $p(x,y) = e^{iwkh(x,y)}$, where $i$ is the imaginary identity, $k=2\pi/\lambda$ in the wave number, and  $w$ is the third element of the key vector. Lastly, $A(\omega_i,\omega_o)$ is a relative, net attenuation term that incorporates the Fresnel term, geometric attenuation and the masking and shadowing effects. More specifically,
\begin{align}
    A(\omega_i,\omega_o) 
    &=R^2(\omega_i,\omega_o) G(\omega_i,\omega_o) / R_0^2 w^2, &\text{for } |w| > w_\epsilon, \nonumber\\
    &= 0 &\text{otherwise.}
\end{align}
Here, $R$ is the Fresnel term, and $R_0$ is the reference Fresnel term for normal incidence under normal viewing, i.e. $R(\omega_i = 0,\omega_o = 0)$. $G$ is the standard geometric term expressed as 
$G(\omega_i,\omega_o) = {(1+\omega_i \omega_o)^2}/{cos\theta_i cos\theta_o}$, 
with $\theta$ as the polar angle for respective direction vectors. $w_\epsilon$ is a threshold (around $10^{-6}$) used to emulate a masking and shadowing function for extreme grazing angles. 

For practical operations, Equation~\ref{eqn:DTFT} can be simplified using Taylor series expansion and Gaussian coherence windowing of the discrete phasor function $p(x,y)$. See \textit{Appendix A} in the \textit{supplemental material} for details. The $n$th Taylor term includes the discrete Fourier transformation (DFT) $\mathscr{F}\{h^n(x,y)\}$. The Gaussian coherence window is parameterized with the standard deviation $\sigma_s$. $N$ is kept arbitrarily large to ensure series convergence. These DFTs are sampled along the $u$ and $v$ dimensions and stored as precomputed lookup tables. 

Now, to generate the ground truth, we first perform spectral sampling for Equation~\ref{eqn:DTFT} at $5$nm resolution for a given set of $(u,v,w)$ tuples. The resulting spectral response is collapsed under spectral integration while scaling it with $D65$ illumination. For the reasons explained later, we default all $A$-terms in Eqn~\ref{eqn:DTFT} to $1$. Resulting map is in CIE-XYZ color space and for the sake of brevity, we refer to it as the \textit{reflectance map}. 

\noindent \paragraph*{Representative Dataset} For our empirical studies, we need a dataset that represents nuanced variations of natural gratings without becoming a pedagogical case or an outlier. Blazed and holographic gratings are regular structures and lack variations. A random initialization or selecting some real-world nanostructural scans can make visual inspections or interpretations tedious and ambiguous. Thus, we construct a synthetic CD profile with realistic random placement of pits and lands to emulate its physical counterpart (see Fig.~\ref{fig:examples}). This reference nano-patch results in one main strong band of rainbow colors spreading across the tracks and several outward streaky blobs that are typical of natural gratings (See a BRDF slice in Fig.~\ref{fig:slice00_2000}). We use the forward modeled reflectance dataset for this synthetic CD as our representative case through most of our ablation studies.   

\subsection{Characterizing Diffractive Reflectance Space}
\label{sec:dataPrep}
There are three main challenges to data-driven modeling of diffractive reflectances: (a) data volume, (b) distributional variations, and (c) range space dynamic variations. Each of these challenges are systematically address in the proposed method, as explained next.
\paragraph*{Data Volume Reduction}
Diffractive reflectances generally span all four polar coordinate dimensions ($\omega_i\equiv(\theta_i, \phi_i);\omega_o\equiv(\theta_o, \phi_o)$) with notable variations.   
For dimension reduction, we exploit the Fourier optical formulation in Equations~\ref{eqn:DTFT},  to factorize diffractive BRDFs into $A$-terms and the remaining part, say, $\mathcal{F}$-terms. The $A$-term still exhibits four dimensional variations but it is computationally inexpensive. $\mathcal{F}$-term depends only on the key vector $(u,v,w) = -\omega_h$. We thus focus on neural representation of this three-dimensional $\mathcal{F}$-term alone. Even in this reduced three-dimensional $(u,v,w)$-space, there are rich variations requiring dense sampling resolution.  \citet{dhillon2014} require a $501\times501\times3\times80$ lookup table space for intermediate results. Even after ignoring the RGB dimensions, this storage requirement is on the order of millions of parameters (around $60$ million in total). Their follow-up work imposes polynomial fitting along the $w$ dimension~\cite{dhillon2016}. They still require around three million parameters. These reference model parameter sizes indicate the need for a much larger training set. We empirically found that $1000$s of samples along both of $u$ and $v$ dimensions and around $11$ samples along the $w$-dimension adequately retain the rich data variations in general. 
These reductions result in practically manageable training timeframes (around $2$ hours per data-set at $73\%$\textemdash$27\%$ split between training and test sets) which can be further reduced as discussed in our ablation studies. 

\paragraph*{Domain Transformation} 
Even with $11$ million sample points in the ground truth, non-linear distribution of the BRDF energy along $u$ and $v$ dimensions can lead to uneven distribution of the errors, resulting in visual artifacts. We thus perform the following nonlinear domain transformation prior to estimating the ground truth. 
\begin{align} (u,v) = \left( 2\text{sign}(u^*) \left(u^*/2\right)^2, 2\text{sign}(v^*) \left(v^*/2\right)^2\right). \end{align}
Here, $u^\ast,v^\ast \in [-2,2]$ represent uniform sampling of their respective intervals. We refer to $(u^\ast, v^\ast)$ as \textit{regular} domain sampling and $(u, v)$ as \textit{simple} domain sampling as long as the $w$ dimension is uniformly sampled along its interval $[-2,0]$. Both regular and simple sampling lead to some data samples that are not physically plausible since sum of two unit vectors, $\omega_h$, cannot have a norm $\omega_h$ greater than $2$. To maximally use the sampling volume, we thus devise another scheme where the $w$ dimension is sampled uniformly along a valid range of $[- sqrt(4- u^2 - v^2), 0]$, which is dependent on the $u$ and $v$. We call this scheme as \textit{simple max} sampling. This still leaves roughly $7\%$ of sample points that are not physically plausible and are set to zero to ignore them. 
With this domain transformation, we improve data learnability of the reflectance map significantly. 

\paragraph*{Range Space Transformations} It is common to apply a range space transformation of $f(x) = log_2(1+x)$ to handle the dynamic range of most reflectance map~\cite{kuznetsov2021NeroMip}. However, this standard range transformation does not address range variational for diffractive reflectances. Histogram analysis of several datasets reveals that though the range space non-linearity has logarithmic characteristics, it is far more compressive at lower values (higher votes in the histogram). We thus devise a bit-plane decompression formulation, 
\begin{align}
    f(x) = 1 + {\log_2 (x)}/{B_\text{max}}\,\,\, \text{where} 
    \label{eqn:logBasic}
\end{align}
$B_\text{max}$ is the floating-point precision, and we refer to it as the \textit{logDivisor}. To avoid range violations, we clamp the transformations to the interval $[0,1]$. From empirical studies, we found the decompression in Eqn.~\ref{eqn:logBasic} to not address low scale values that show systematic errors at high exposure for rendering. We thus devise two other transformations for examinations.
\begin{align}
    g(x) &= \text{pow}(x, n), \text{ with $n$ as a \textit{power} parameter, and} \\
    f2(x) &= \text{pow}(f(x), n). \label{eqn:logPower}
\end{align}

In our empirical studies, we found the transformation in Equation~\ref{eqn:logPower} to provide the best reconstruction, in general.

\subsection{Neural Representation}
\begin{figure*}[t!]
    \centering
    \includegraphics[width=0.85\linewidth]{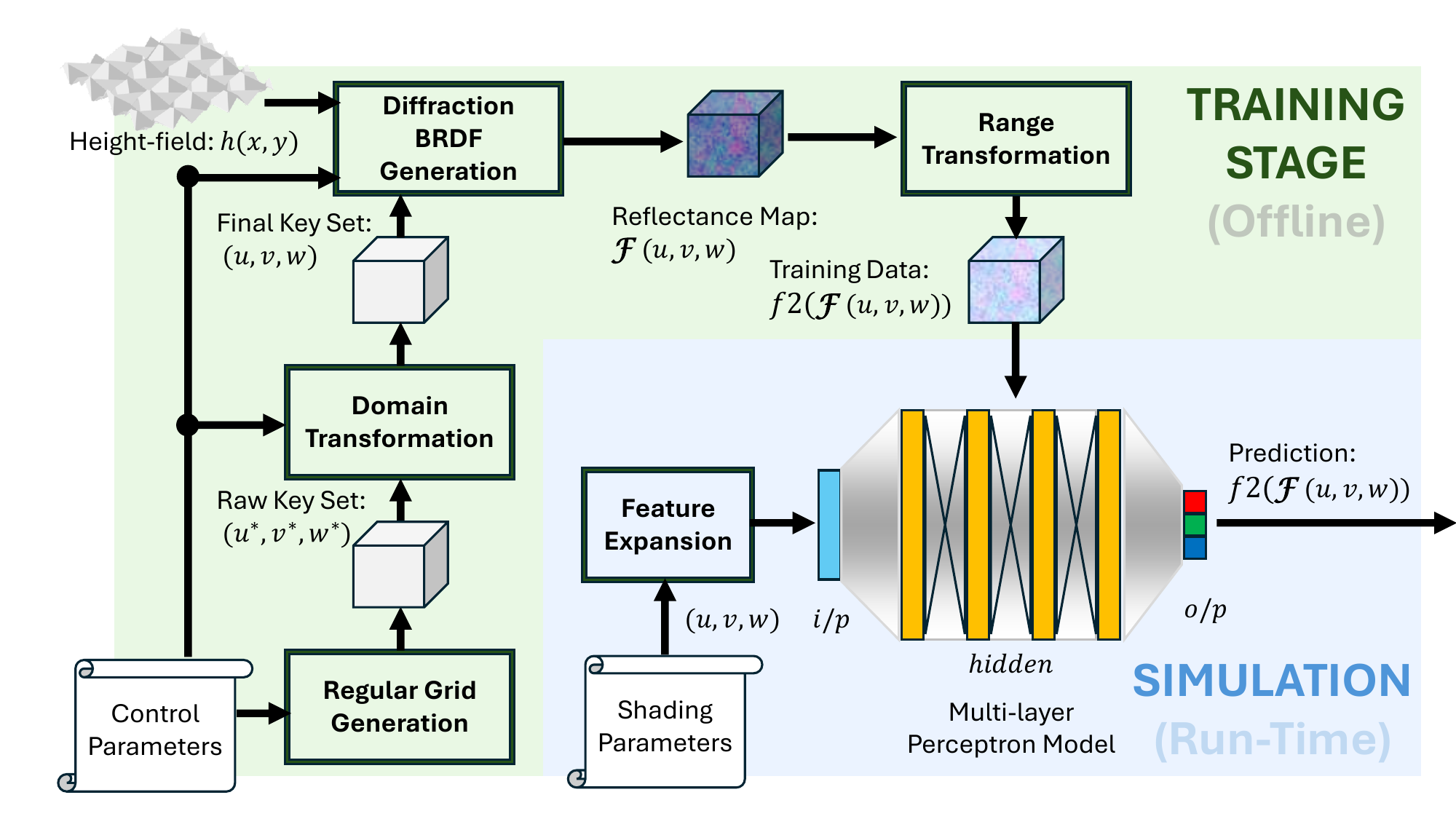}
    \caption{Operational workflow for the proposed method. }
    \label{fig:Setup}
\end{figure*}
For effective neural representation, it is critical to design, examine and fine-tune the model networks. For operational efficiency, we chose to work with multi-layer perceptrons. In this section, we first present the operational workflow for the proposed MLP-based method.  \paragraph*{Workflow}
Figure~\ref{fig:Setup} illustrates our workflow in operation for a single dataset belonging to one given height-field profile. It consists of two stages. The first stage is offline, and it involves setting up the sampling scheme, data generation, range transformation, data segregation into training and test sets and actual training of the neural network. The key blocks for this stage are explained in Section~\ref{sec:dataPrep}. The second stage for runtime simulations sequences through shading context preparation, input feature expansion, driving the network and assimilating the MLP predicted XYZ color vector into the frame buffer. The shading context preparation task produces the following shader parameters: (i) the light direction $\omega_i$, (ii) view direction $\omega_v$, (iii) surface normal for each pixel and uploads the MLP network as a texture. As the first step for the runtime stage of the proposed method, we compute the input feature vector using $\omega_i$ and $\omega_v$. 

\paragraph*{Position Encoding and Feature Expansion}
Sztrajman et al.~\cite{timWeyrich2021} use just the Rusinkiewicz re-parameterization of the $\omega_i$ and $\omega_v$ direction vectors to model reflectance maps (BRDFs) with MLPs. Their target BRDF dataset lacks high-frequency features, and thus, a simple six-element input feature vector proves to be effective. Tancik et al.~\cite{tancik2020fourier} demonstrate that such an approach does not reconstruct sharp local variations in image, 3D shape of other data regression problems. They propose to encode the frequency space characteristics into the positional vector before passing it on as the input feature to query the MLP. This encoding of the position vector $\mathbf{v} \in [0,1)^d$ space with the transformation $\mathcal{B}$  that expands in to a hyperspace $[-1,1]^{2m}$ can be formulated as~\cite{tancik2020fourier}:
\begin{align}
\mathcal{B}(\mathbf{v}) &= \left[  a_1 \cos(2\pi \mathbf{v}^T\mathbf{b}_1), a_1 \sin(2\pi \mathbf{v}^T\mathbf{b}_1), ...,\right. \nonumber\\
  \quad\quad &\quad\quad \left. a_m \cos(2\pi \mathbf{v}^T\mathbf{b}_m), a_m \sin(2\pi \mathbf{v}^T\mathbf{b}_m) \right] ^T
  \label{eqn:posEncode1}
\end{align}
In the equation above, $\{\mathbf{b}_i\}$ represent $m$ most prominent frequencies and $\{a_i\}$ represent their prominence measures respectively, for a given domain. This positional encoding scheme requires careful selection of Fourier basis frequencies that are commonly present in a given dataset. Simply put, determining vectors $\mathbf{b}_j$s and scalars $a_j$s establishes $m$ stationary characteristics of the data domain. 
In the context of image compression, ~\citet{strumpler2022implicit} introduce a parametric set of basis frequencies, all of equal prominence, by adapting the positional encoding in Equation~\ref{eqn:posEncode1} to the following form:
\begin{align}
\mathcal{B}(\mathbf{v}) = \left[\mathbf{v}^T , \sin( s^0\pi \mathbf{v}^T), \cos(s^0 \pi\mathbf{v}^T), ...,\right.  \nonumber \\
\left. \sin(s^m \pi\mathbf{v}^T), \cos( s^m\pi \mathbf{v}^T) \right] ^T.
\label{eqn:posEncodin2}
\end{align}

Note that their position vector is two-dimensional and $\mathbf{v} = (v1,v2) \in (-1,1)^2$. Also, they augment the original position vector to the encoded one. Lastly, $s$ is the parameter that one can tune in to modulate the geometric series of the basis frequencies.

Diffractive reflectance data, which spans three dimensions $\mathbf{v} = (u,v,w)$, often exhibits high-frequency details and can benefit from positional encoding. However, Equations~\ref{eqn:posEncode1} or~\ref{eqn:posEncodin2} cannot be directly applied to the domain defined with the key vector $(u,v,w)$. While $u,v\in[-2,2]$ can be simply scaled and offset to suit above equations (due to periodicity of DFTs), such scaling is not a trivial matter in general. To appreciate the subtlety, consider the case for $w\in[-2,0]$. We need to understand the implications of such domain transformation. By transforming the range for $w$ to $[0,1)$ in applying Equation~\ref{eqn:posEncode1} or to the range $(-1,1)$ in applying Equation~\ref{eqn:posEncodin2} we implicitly impose periodicity across the interval with the use of the Fourier basis. This is not the key characteristic of the diffractive reflectance maps and we found a trivial application of position encoding to the $w$ space to struggle with the learning tasks. 
We thus remap $w$ to the range $[0,1]$ and use the positional encoding scheme in Equation~\ref{eqn:posEncodin2}. This avoids the imposition of the periodicity, and we are able to use positional encoding to improve our modeling accuracy. 

More importantly, we first expand the position vector to $\mathbf{v} = (u,v,u+v,v-v, w)$ before substituting it in Equation~\ref{eqn:posEncodin2}. We found this explicit phase rotations in the Fourier space to significantly reduce errors. Also, we introduce the encoding scheme to set the value of $s$ based on the understanding of Nyquist sampling. We first fix the $m$ number of basis vectors to use and then set $s$ such that $s^m$ corresponds to half the sampling frequency, which is the highest frequency in the discrete data. This principled approach allows for systematic resizing of the input feature vector to adapt for network resizing.

\paragraph*{Network Design}
Size, shape and depth of the hidden layer block in an MLP model, each of these factors impact the learning capacity. Based on our experiments, we propose to use a funneling mechanism in structuring hidden layers for  MLPs modeling diffractive reflectances. For funneling, we make the layer immediately after the inputs to be the largest. Each consecutive layer is reduced in size by a constant scale factor. Funneling reduces overfitting and  balances learning competition by emulating simulated annealing under exponentially decaying network size. There are five design factors counter-playing each other: (i) total network size, (ii) network depth, (iii) start size, (iv) end size, and (v) the funneling scale factor. We generally found the funneling scale ratio to the reciprocal of the golden ratio (1.618) to produce consistent result across the frequency spectrum. Also, we found that having 1.5 times to twice the input nodes for the first hidden layer allow for adequate combinations across the frequency basis vectors.  
Based on these design principles, we experimented with several network sizes and found this funneling mechanism to produce consistent results with stable correlation between the network size and the reconstruction accuracy.  

\paragraph*{Training} Generally, we use around $73\%$ ($8$ out of $11$ $w$-slices) of the ground truth as the training data and remaining as the test data. We tried two segregation strategies with similar performances. In the first strategy, we randomly select training samples from anywhere within the data volume of $1001\times1001\times11$ samples across $(u,v,w)$ dimensions respectively. In the second and the default strategy, we pick three slices along the $w$ dimension (Slice no. 4, 7 and 10) for testing. This allows use to examine the use the SSIM measure as well as visual inspections to reason appropriately about the network accuracy. Lastly, for the most difficult cases of complex fabrications, we use all the data in training, akin to the image compression methods~\cite{catania2023nif}. For such cases, we additionally examine the model efficacy by analyzing BRDF slices with fixed incident light direction/s $\omega_i$. For such slices, pixels rarely coincide with the samples from the training data and thus work well to examine the model predictions. 

\section{IMPLEMENTATION DETAILS}
\label{sec:Implementation}

\paragraph*{Datasets}
We use three group types of datasets: \textit{basic}, \textit{real-world scans}, and \textit{complex designs}. We construct the height-fields in the \textit{basic} dataset using common references. This group consists of: (i) a blazed grating, (ii)  a statistical CD patch, and (iii) a random height-field with Gaussian-windowing. 
All these height-field patches are shown in Figure~\ref{fig:examples}. 
The maximum height of the blazed grating is $250nm$, it has a periodicity of $2.5\mu m$ and a patch size of $100\times100 \mu m^2$. A blazed grating has a high diffraction efficiency it diffracts most of the incident light energy into one single mode of diffraction. This results in a reflectance map with very little distribution and a strong anisotropic spread. This can help in examining any learning biases that will usually manifest as easily noticeable visual artifacts.

Synthetic CD emulates grooved tracks on a read-only compact disk (CD-R) to the real-life scales. Tracks are $1.6\mu m$ apart and the patch size is $65\times65 \mu m^2$. The CD nanostructure has pits and lands that are randomly placed along the track but with discretized spacing. The pits can only span few bits contiguously. This leads to very interesting, varied ensemble of rainbow streaks arising from the first mode of diffraction along the track directions. Also, the primary diffraction across the tracks is sharp, vivid and, like blazed grating assists visual inspection. We thus use it as our \textit{reference dataset} in most of the ablation studies. For examining extreme complex spreads, we also devise a random height-field with maximum elevation of $1\mu m$ and patch size to match most of the others, i.e. $65\times65 \mu m^2$. However, due to rectangular windowing of the patch, it results in visual artifacts due to frequency aliasing. We thus impose a Gaussian window on the random height-field to elevate such aliasing issues.

For working with actual microscopic scans of real-world nanostructures, we obtained two snakeskin scans by the courtesy of \citet{dhillon2014}. One belongs to a corn snake (Elaphe) and the second one is for a sunbeam snake (Xeno). Both the patches are $65\times65 \mu m^2$, and further details are available in their paper. We also downloaded a scan of scratches on the tip of a ballpen, shared courteously by NanoSurf Corporation~\cite{nanoSurf}. Firstly, we pre-process this scan image to upsample its required resolution using an online, high-quality API~\cite{deepAI}. Next, we rescale and crop it to $80\times80\mu m^2$. Also, since the scratches are not quasi-periodic, we impose a Gaussian damping window for alleviating aliasing effects. The resulting patch with maximum elevation of $873nm$ is shown in Figure~\ref{fig:examples}. Lastly, we also examine our model with two highly \textit{complex design} constructs recently published and shared by the courtesy of \citet{yu2023}, namely \textit{Corner Cubes} and \textit{Spherical Pits}.    

\paragraph{Ground-Truth Generation Details} We use the reference method presented in Section~\ref{sec:groundtruth} to generate the ground truth datasets for learning. \citet{yu2023} have noted the limitations of producing correct reflectances for Corner Cubes using Kirchhoff's integrals. We found the primary limitation relates to its maximum height. For uniformity's sake and to go beyond the narrow coherence waist used in their work, we pre-process the original Corner Cube patch of $24\times24 \mu m^2$ through mirrored repetitions and height down-scaling for a maximum elevation of $1.5 \mu m$. The resulting patch of $72\times72 \mu m^2$ is shown in Figure~\ref{fig:examples}. It produces reflectances qualitatively similar to those in the original paper~\cite{yu2023} when processed by our reference Fourier shader. Similarly, the reference patch for Spherical Pits has very high resolution to process with our reference shader. We thus downsample it to the resolution of $320 \times 320$ pixels, which is shown by \citet{yu2023} to produce subjectively similar results. Lastly, we emphasize that their original method is prohibitively expensive to operate on our platform at the coherence length and wavelength sampling comparative to our other patches. We thus re-purpose their height-field models through our reference Fourier shader which produces qualitatively similar reflectance maps.

\paragraph*{Network Training and Evaluation Details} 
At a more technical level, we employ the \textit{LogCosh} loss function, as it behaves like Mean Squared Error (MSE) for smaller values and Mean Absolute Error (MAE) for larger values, ensuring stability across varied error magnitudes. Also, a dynamic learning rate scheduler is utilized to adapt the learning rate during training, improving convergence. Training was conducted on a machine equipped with two RTX 3090 GPUs, significantly reducing computation time.

\paragraph*{Default Configuration}

Based on our empirical studies, we set our default configuration to produce the best qualitative results. In this configuration, the network has a size of ($163$, $464$, $288$, $176$, $108$, $68$, $40$, $24$, $16$, $3$) including both input and output layers. This architecture expands the low-dimensional input using sinusoidal encodings, with nineteen frequencies along the $u$ and $v$ dimensions, and four along the $w$ dimension. While this model produces the best result, we observed that consistent quality with smaller models. We use the range transformation as given by Eqns.~\ref{eqn:posEncodin2} \&~\ref{eqn:logPower}, with $n=8$ and the \textit{logDivisor} ($B_\text{max}$) set to $48$. We use the \textit{simple max} sampling for ground truth generation, by default. 

\paragraph*{Evaluation Metrics}
In order to evaluate the model's performance, we used the Peak Signal-to-Noise Ratio (PSNR), \FLIP, and 
Structural Similarity Index Measure (SSIM). 
PSNR in the XYZ color space provides an evaluation of the predicted image in a device-independent color space, thus making it more reliable in quantifying learning losses in the raw data space. 
\FLIP \cite{andersson2020flip} is a difference evaluator that quantitatively characterizes the perceptual differences that are observable while flipping two images. Having a smaller mean \FLIP value is considered better.
For SSIM, we use YCbCr color space as recommended for evaluating the renderings and videos~\cite{nilsson2020understandingssim}. We also include SSIMs in sRGB color space for direct subjective assessments.

\paragraph*{Rendering Exposure} The reference method for generating the ground truth (see Sec.~\ref{sec:groundtruth}) uses a relative attenuation factor. This is akin to fixing the exposure for imaging. At one unit of this relative exposure (say 1RU), the structural colors due to diffraction are very weak. We thus generally use a relative exposure from $1000$\textendash$5000$ RUs. This approach is similar to the one used by \citet{dhillon2014}.

\section{RESULTS}
\label{sec:Results}
\begin{figure}[t!]
\small

\centering
\includegraphics[width=\linewidth]{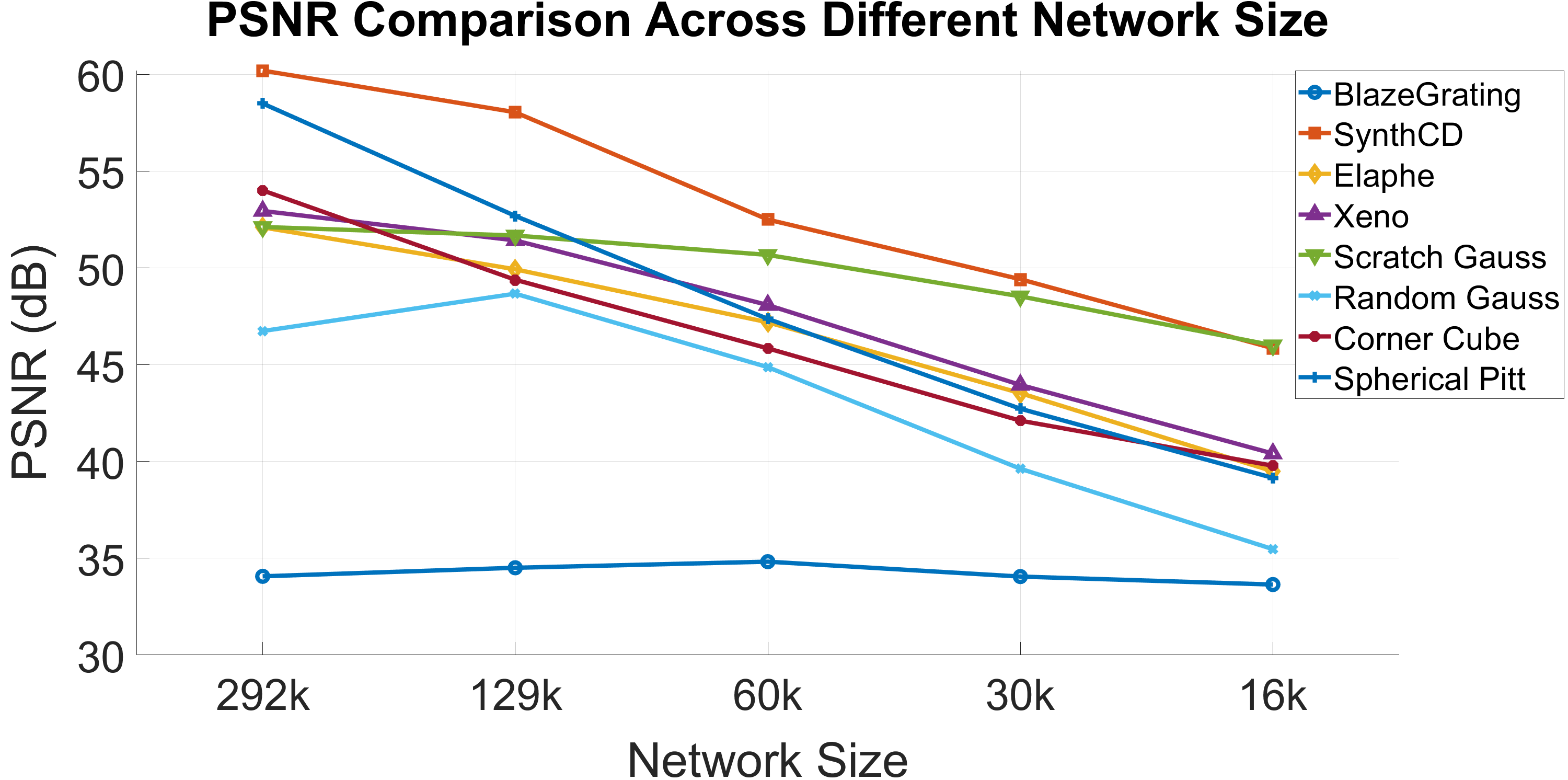}  \\[0.5ex]
a. PSNR Plot \\[3.5ex]

\includegraphics[width=\linewidth]{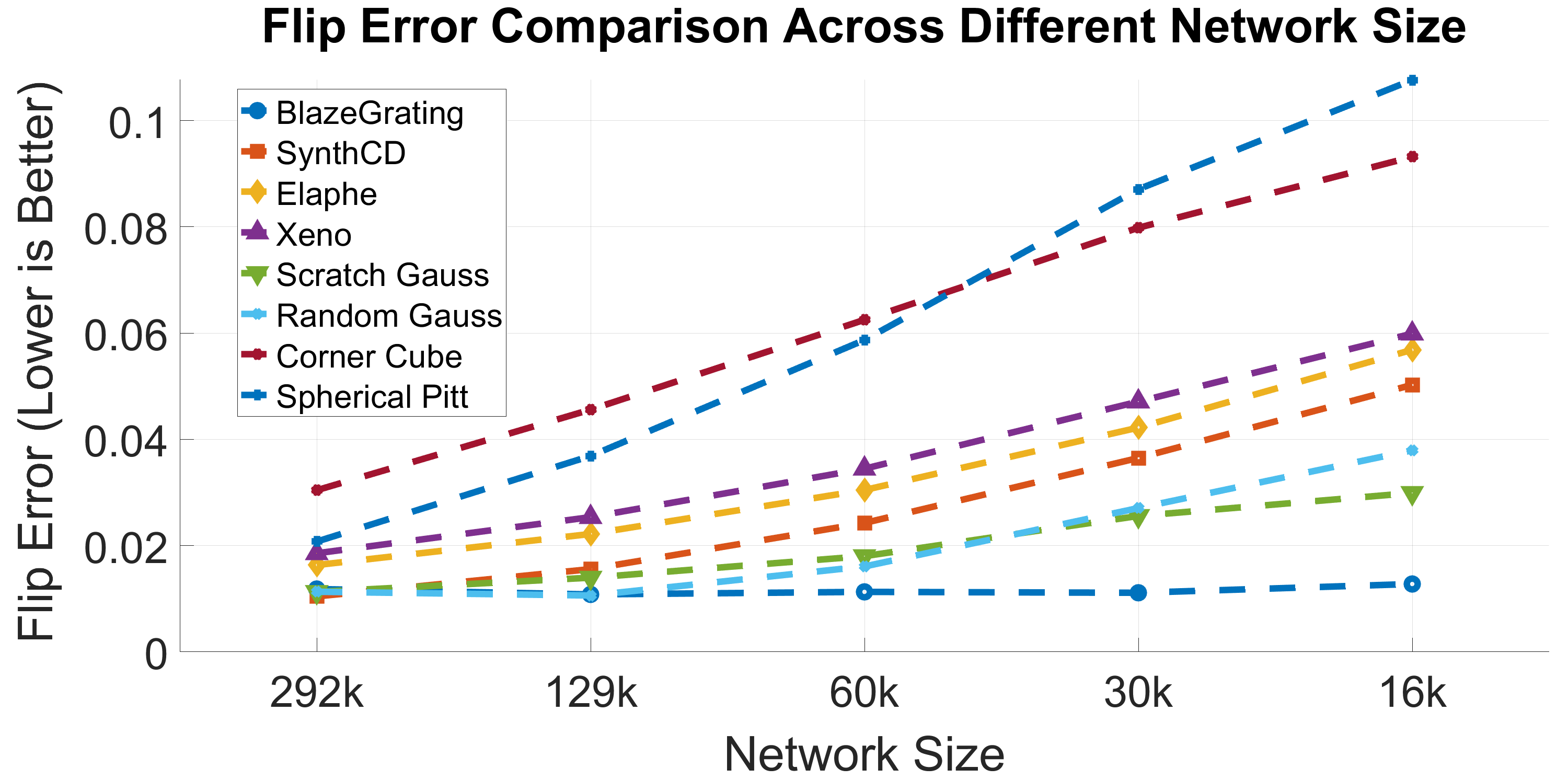} \\[0.5ex]
b. Flip Plot\\[3.5ex]

\includegraphics[width=\linewidth]{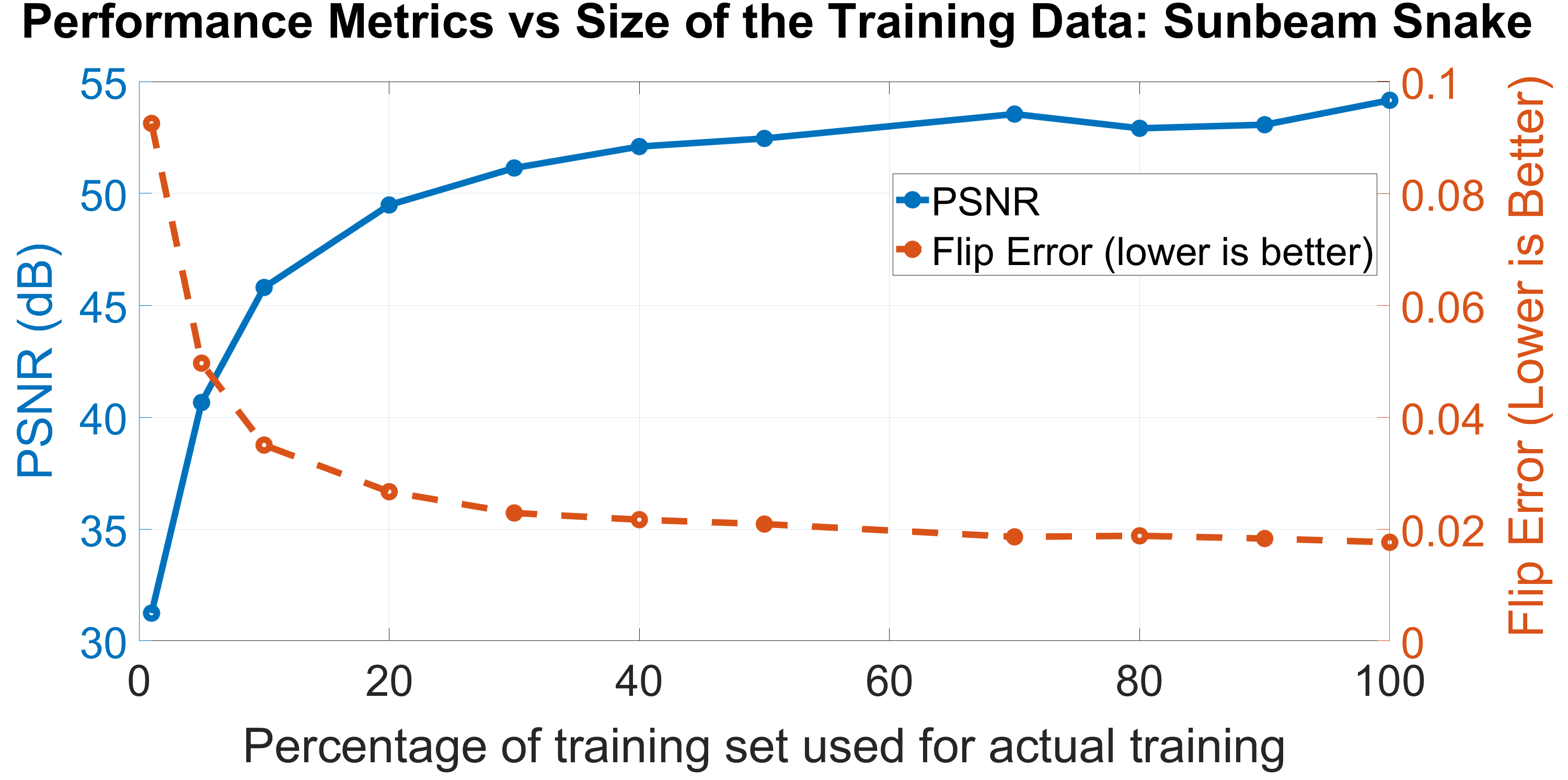} \\[0.5ex]
c. Flip \& PSNR at different training samples.\\
\caption{Comparison of performance metrics and training set statistics across various configurations at $\times2000$ exposure on Slice 10. (a) PSNR plots for different synthetic and complex datasets. As network size decreases, PSNR generally drops—except for the Blaze Grating dataset—indicating that larger networks are more effective at learning complex BRDFs. (b) Flip error corresponding to the configurations in (a). The trend reinforces that larger networks yield better performance across datasets. (c) Network performance on test Slice 10, shown via PSNR and Flip error, as a function of the percentage of training data used. This highlights how training set size influences learning ability.}
\label{fig:PsnrFlip}
\end{figure}

\begin{figure*}[ht!]
    \centering
    \begin{tabular}{r@{\hspace{1mm}}c@{}c@{}c@{}c@{}c@{}c}
    \hfill{} Ground Truth \hfill{} & 10\% & 20\% & 30\% & 50\% & 70\% & 100\%\\
    \includegraphics[width=0.14\linewidth]{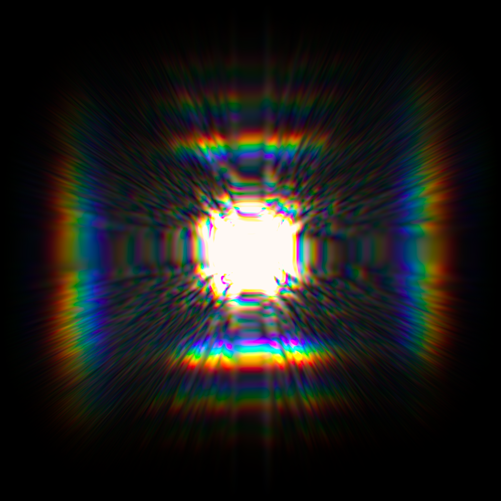} &
    \includegraphics[width=0.14\linewidth]{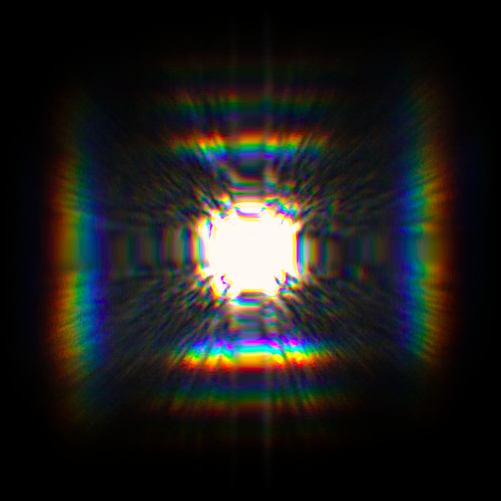} &
    \includegraphics[width=0.14\linewidth]{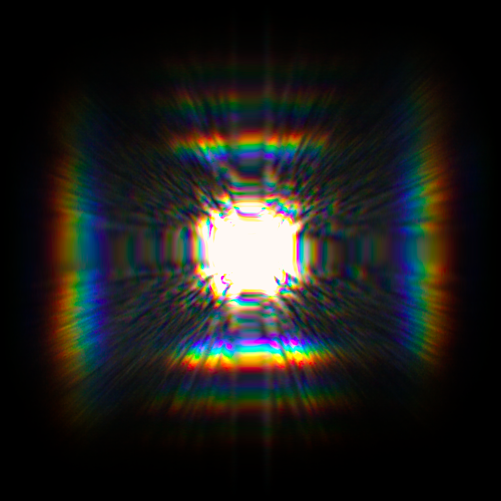} &
    \includegraphics[width=0.14\linewidth]{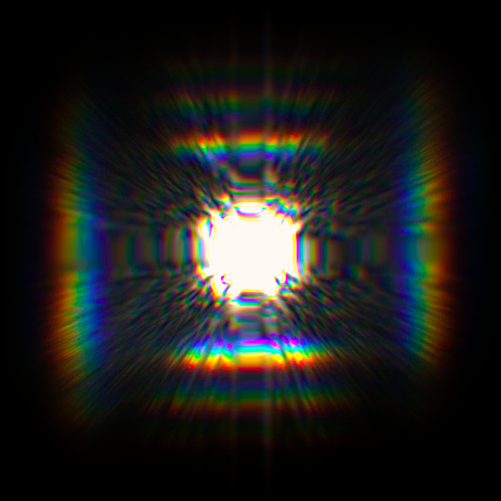} &
    \includegraphics[width=0.14\linewidth]{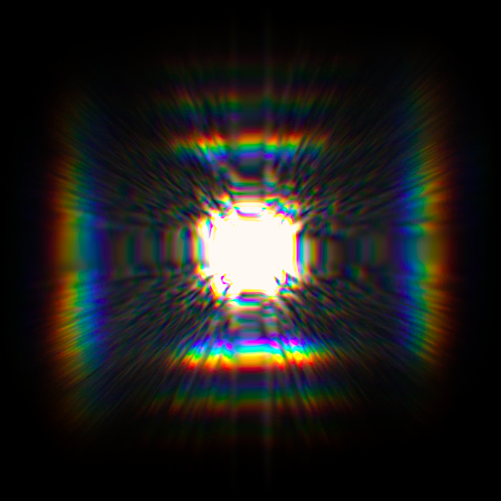} &
    \includegraphics[width=0.14\linewidth]{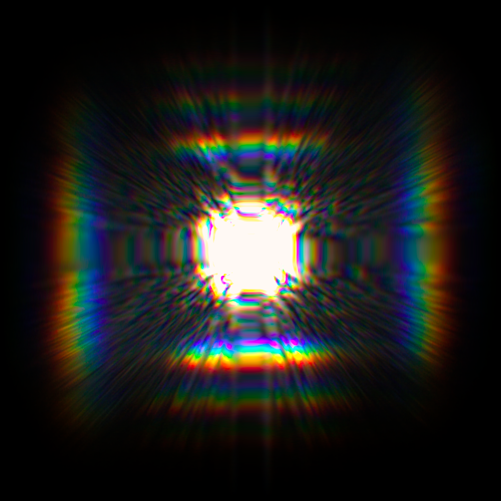} &
    \includegraphics[width=0.14\linewidth]{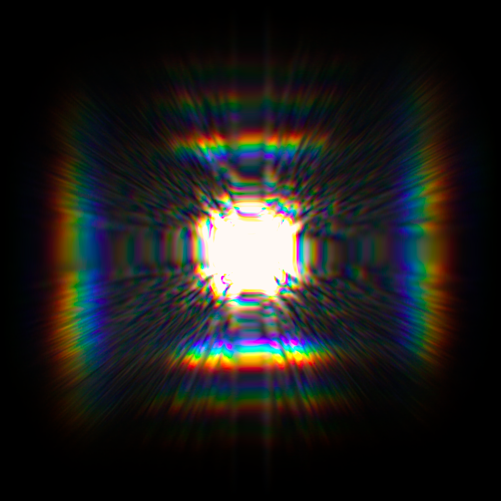} \\
    
    \rotatebox{90}{Flip Error Map} & 
    \includegraphics[width=0.14\linewidth]{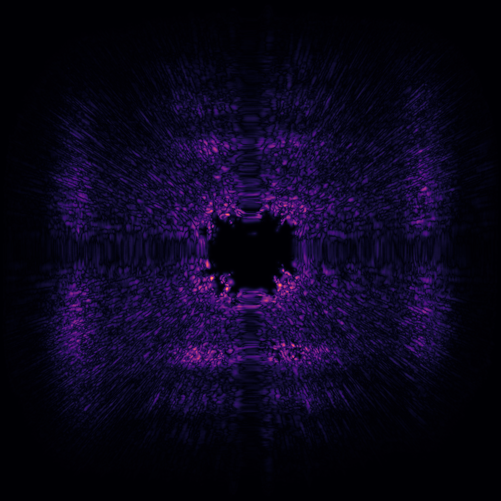} &
    \includegraphics[width=0.14\linewidth]{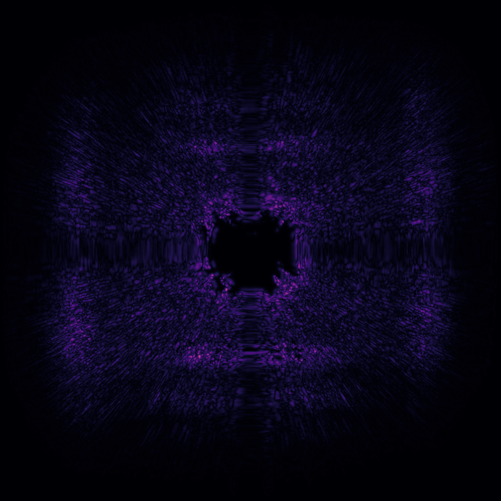} &
    \includegraphics[width=0.14\linewidth]{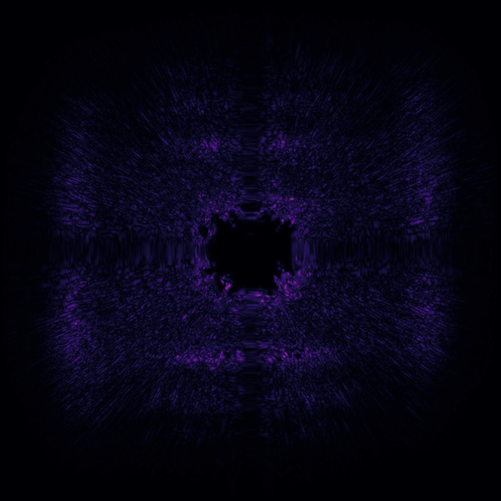}&
    \includegraphics[width=0.14\linewidth]{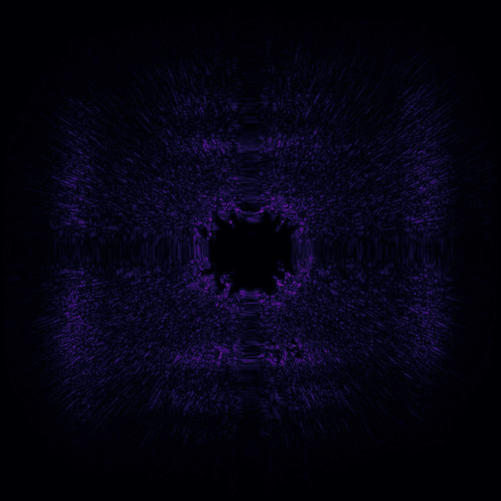}&
    \includegraphics[width=0.14\linewidth]{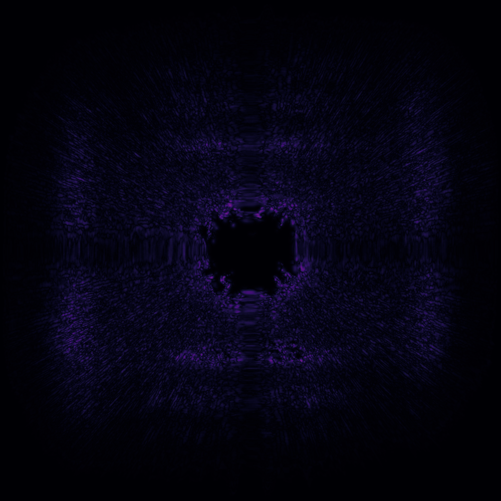}&
    \includegraphics[width=0.14\linewidth]{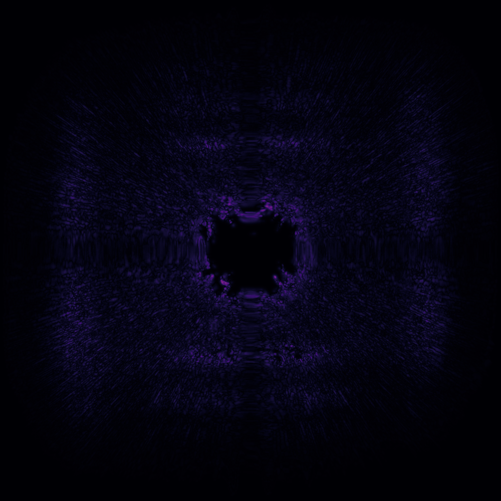} \\
    $\mu$ & $0.0556$ & $0.043$ & $0.036$ & $0.034$ & $0.03$ & $0.028$\\     
    \end{tabular}

    \caption{Network predictions on Sunbeam Snake test BRDF data (Slice 10) at $\times 5000$ exposure, trained with varying sampling rates. Of the 11 total slices, slices 4, 7, and 10 serve as the test set, while the others are used for training. $\mu$ denotes the mean \FLIP Error.} 
    \label{fig:wwFlip}
\end{figure*}

We first examine our method for $\mathcal{F}$-terms (see Section~\ref{sec:groundtruth}) with the default configuration against the test data for the reference Synthetic CD dataset at an exposure of $2000$RU. The PSNR for slices $4$,$7$ and $10$ under \textit{simple max} sampling at $(u,v)$ resolution of $1001\times1001$ are ($56,690, 62.006,60.2$) dB respectively. Similarly, the mean SSIM in sRGB color space for these slices are ($0.9987,0.9984,0.9974$) and the \FLIP scores are ($0.0082,0.0081, 0.0105$) respectively. We also performed this test for other datasets in the \textit{basic} and \textit{real-world scans} groups, which produce similar results. Fig.~\ref{fig:PsnrFlip} (a) and (b) plot PSNR and \FLIP statistics for all these cases along at different network sizes. Resulting error metric statistics confirm high-quality reconstruction by our method, in general. For visual examination at more adverse exposure, we depict the \FLIP error map for the Xeno dataset at $5000$RU (see Fig.~\ref{fig:wwFlip} (last column)). Furthermore, we also examine the robustness of our method when subject to  training data scarcity. Fig.~\ref{fig:PsnrFlip}(c) \& Fig.~\ref{fig:wwFlip} demonstrate with the Xeno case that decreasing the training data set to $30\%$ does not result in any notable differences in the final results. 

\begin{figure*}[t!]
\small
    \centering
    \begin{tabular}{c@{\hspace{1mm}}c@{}c@{}c@{}c@{}c@{}c@{}c}
    
    \rotatebox{90}{Actual}&
    \includegraphics[width=0.14\linewidth]{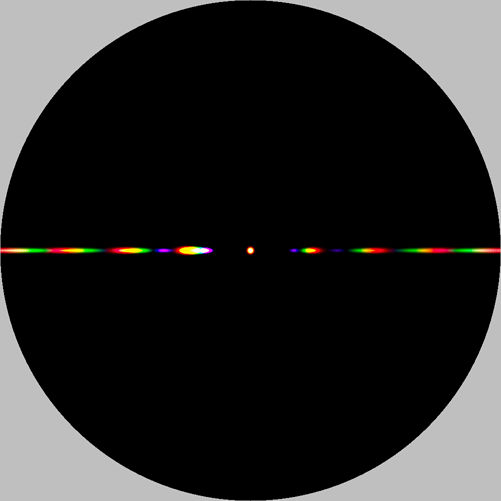} &
    \includegraphics[width=0.14\linewidth]{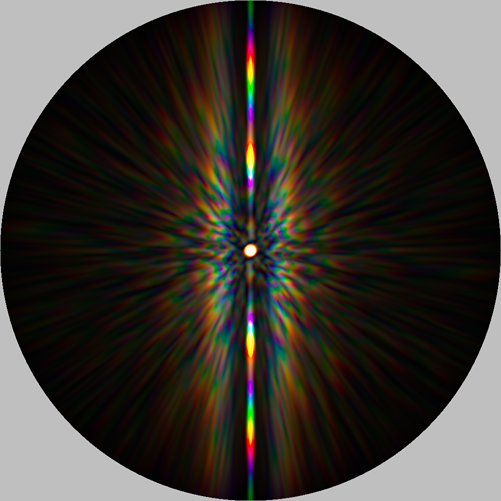} &
    \includegraphics[width=0.14\linewidth]{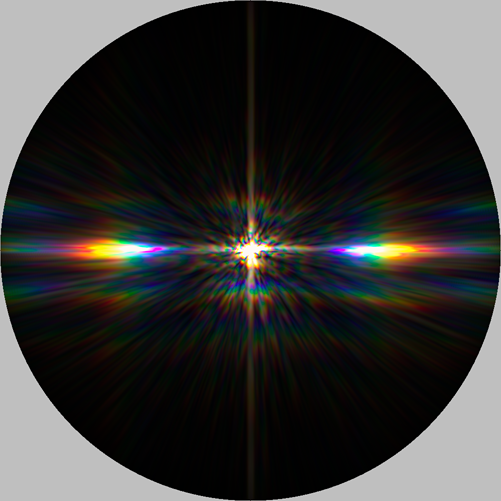} &
    \includegraphics[width=0.14\linewidth]{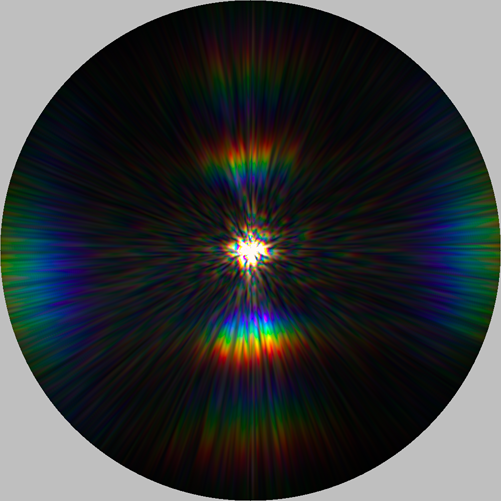} &
    \includegraphics[width=0.14\linewidth]{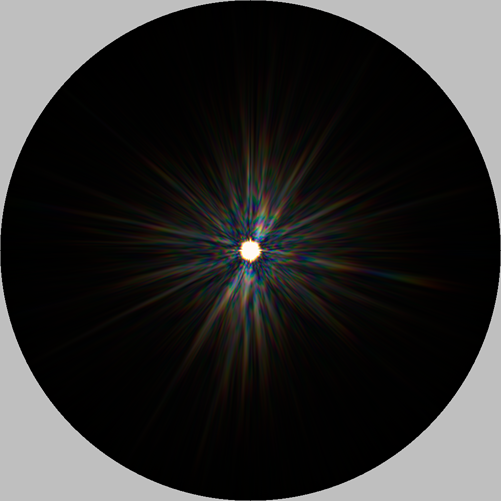} &
    \includegraphics[width=0.14\linewidth]{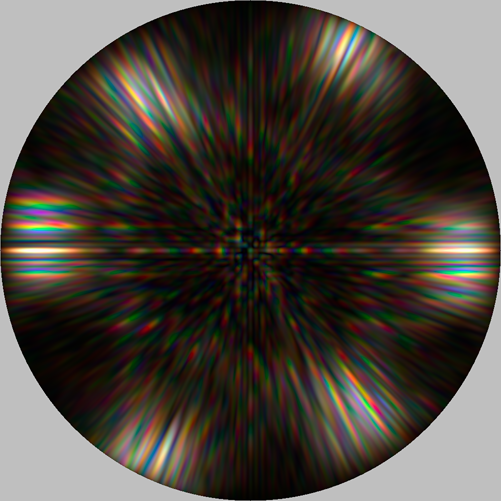} &
    \includegraphics[width=0.14\linewidth]{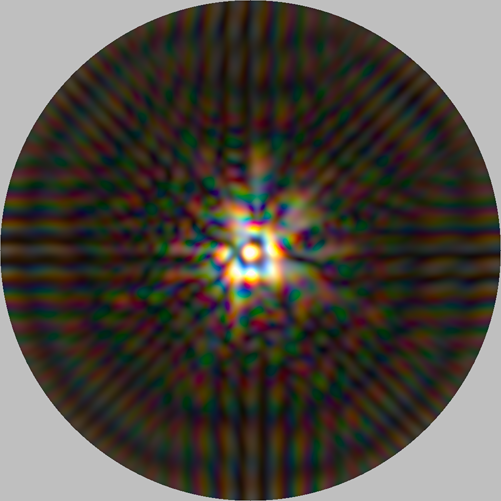} \\
    
    \rotatebox{90}{Predicted}&
    \includegraphics[width=0.14\linewidth]{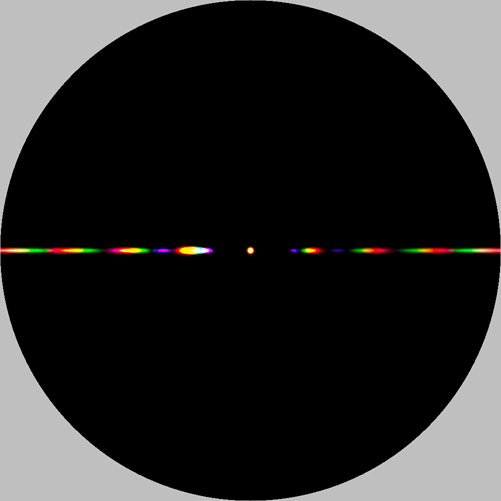} &
    \includegraphics[width=0.14\linewidth]{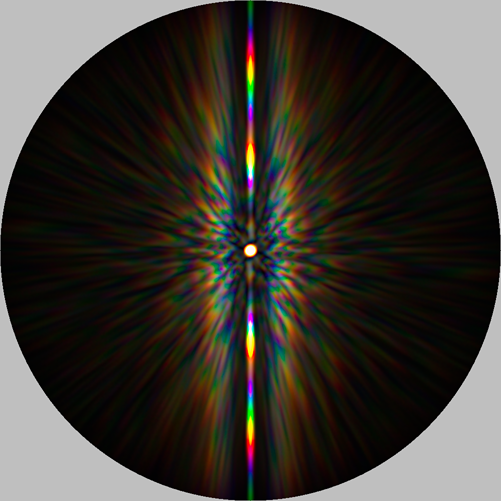} &
    \includegraphics[width=0.14\linewidth]{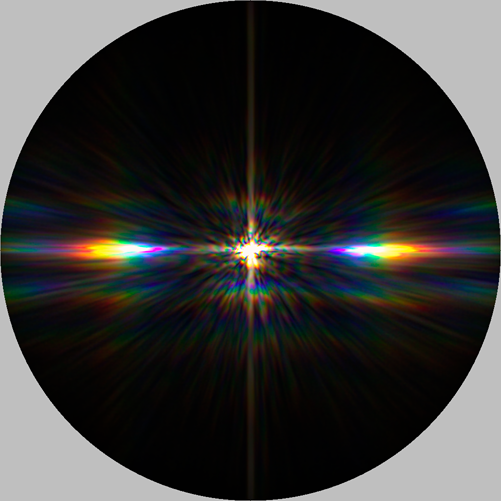} &
    \includegraphics[width=0.14\linewidth]{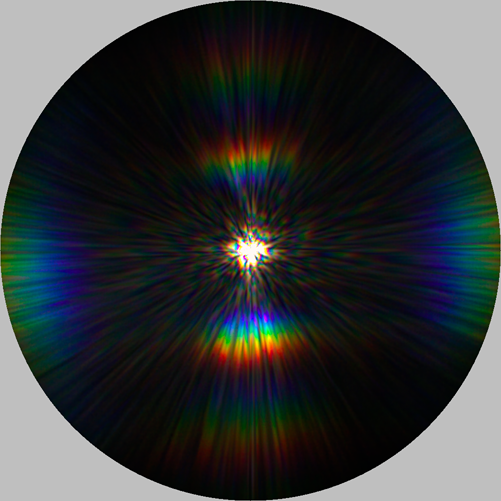} &
    \includegraphics[width=0.14\linewidth]{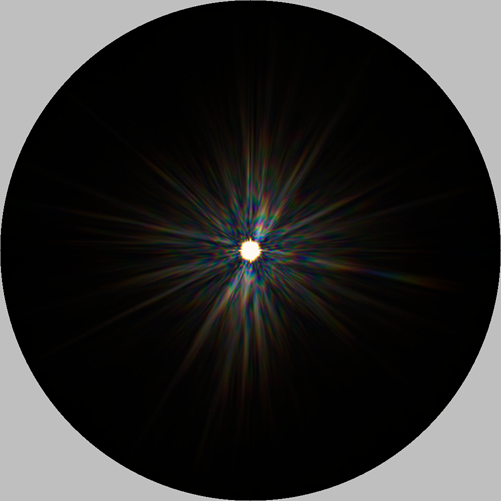} &
    \includegraphics[width=0.14\linewidth]{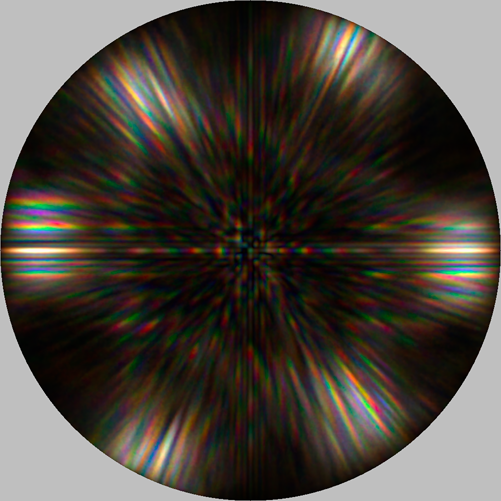} &
    \includegraphics[width=0.14\linewidth]{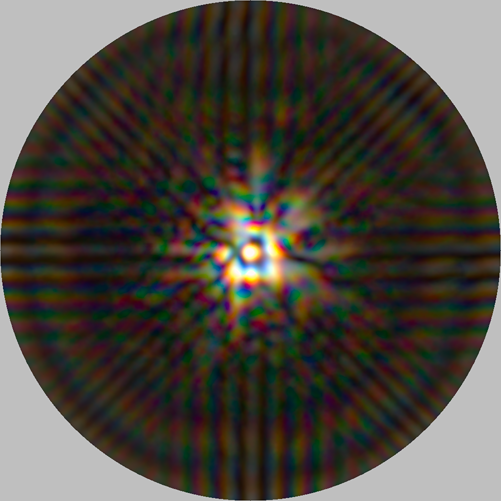} \\
    
    \rotatebox{90}{Flip Error Map} &
    \includegraphics[width=0.14\linewidth]{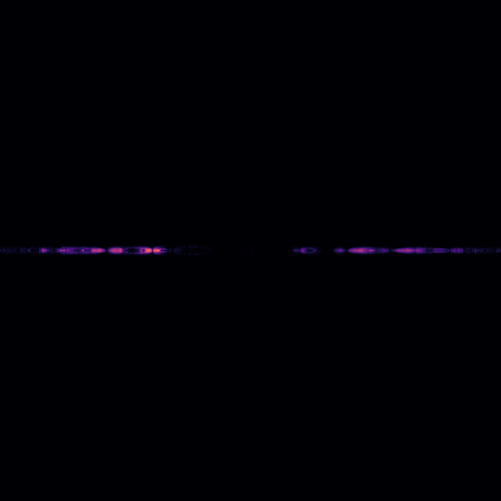} &
    \includegraphics[width=0.14\linewidth]{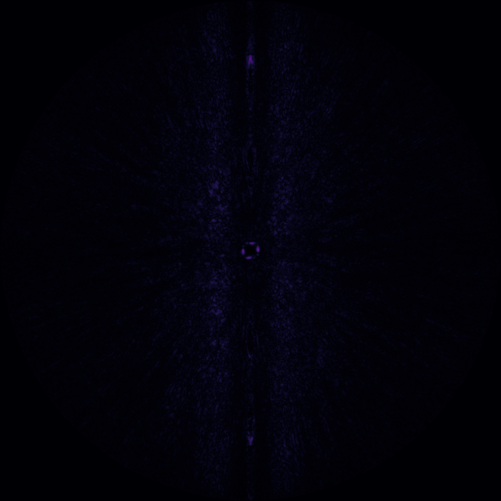} &
    \includegraphics[width=0.14\linewidth]{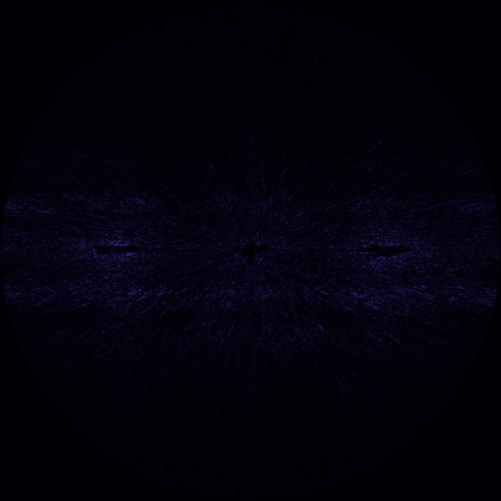} &
    \includegraphics[width=0.14\linewidth]{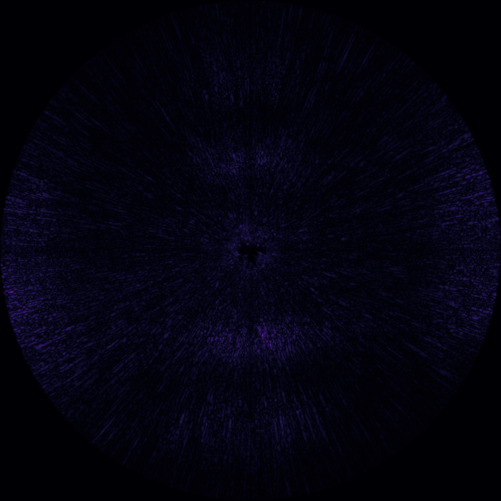} &
    \includegraphics[width=0.14\linewidth]{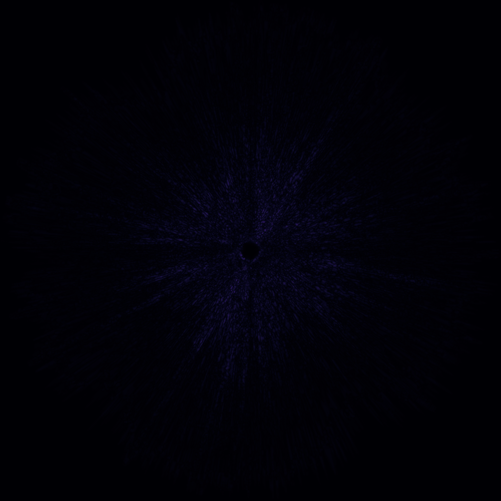} &
    \includegraphics[width=0.14\linewidth]{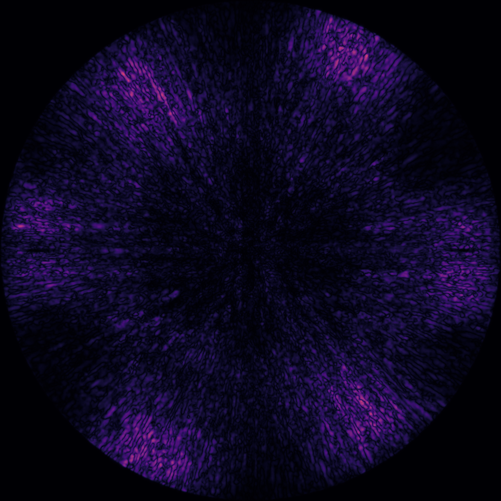} &
    \includegraphics[width=0.14\linewidth]{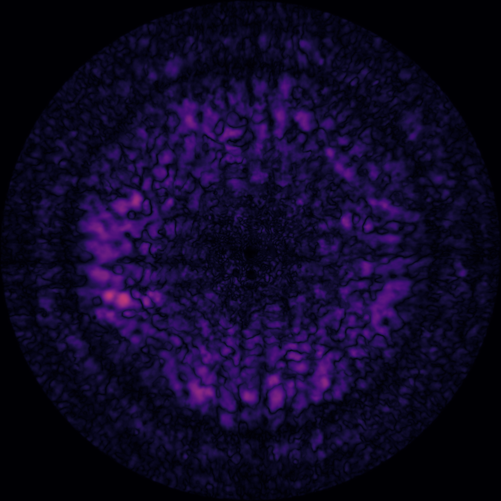} \\
    $\mu$ & $ 0.00212 $ & $0.0178 $ & $ 0.0221 $ & $0.0318$ & $0.0123 $ & $0.0751$ & $0.0878$ \\  
    & Blazed Grating & Synthetic CD & Cornsnake & Sunbeam Snake & Ballpen Tip Scratch & Corner Cubes & Spherical Pits\\
    \end{tabular}

    \caption{BRDF slices for (\(\theta\), \(\phi\)) : \((0, 0)\) at the exposure \(\times 2000\). $\mu$  represents average Flip error.}
    \label{fig:slice00_2000}
\end{figure*}

Next, we validate our method against BRDF Slice renderings. A BRDF slice is generated by fixing the incident light direction and projecting all view directions from the hemisphere onto the base plane. The color at the projected location represents the fully evaluated value of the reflectance map, including the Fresnel effects. The $(u,v,w)$ rarely coincide with the training dataset sample locations and thus rigorously validate the model. 

\begin{figure*}[t!]
\small

    \centering
    \begin{tabular}{@{}c@{\hspace{1mm}}c@{}c@{}c@{}c@{}c@{}c@{}c}
    
    \rotatebox{90}{Actual}&
    \includegraphics[width=0.14\linewidth]{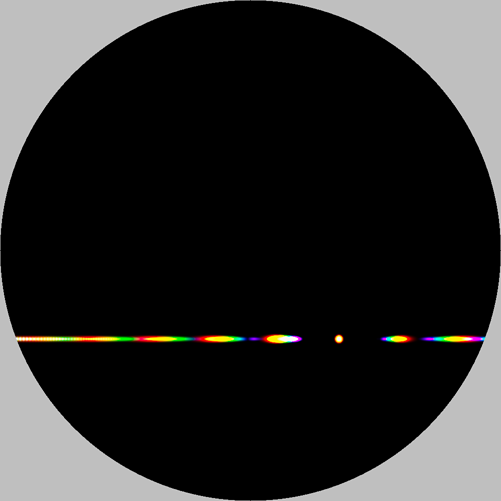} &
    \includegraphics[width=0.14\linewidth]{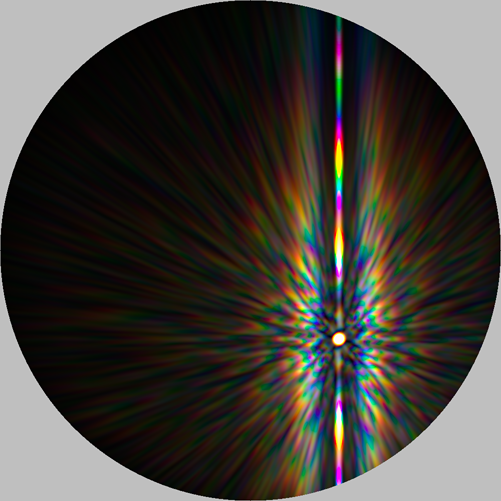} &
    \includegraphics[width=0.14\linewidth]{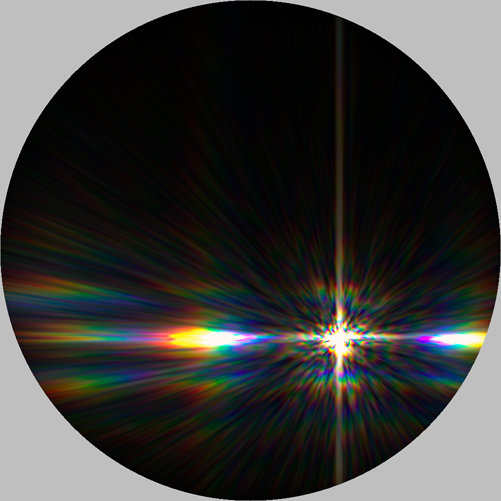} &
    \includegraphics[width=0.14\linewidth]{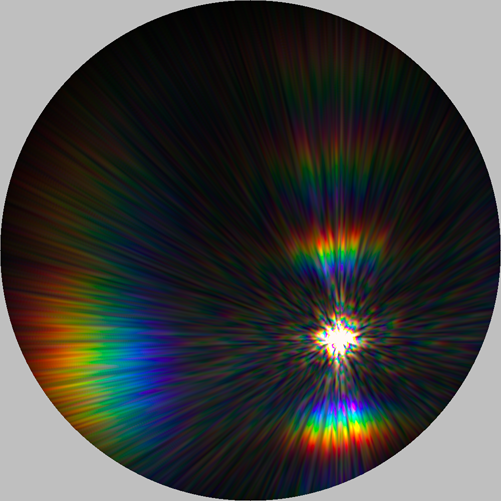} &
    \includegraphics[width=0.14\linewidth]{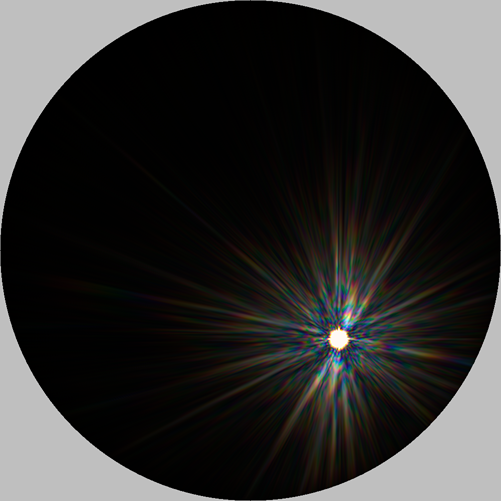} &
    \includegraphics[width=0.14\linewidth]{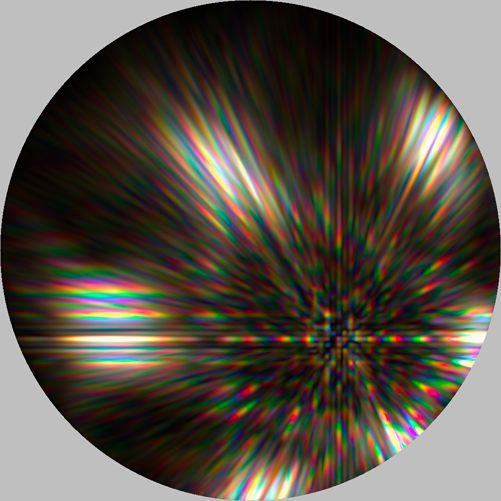} &
    \includegraphics[width=0.14\linewidth]{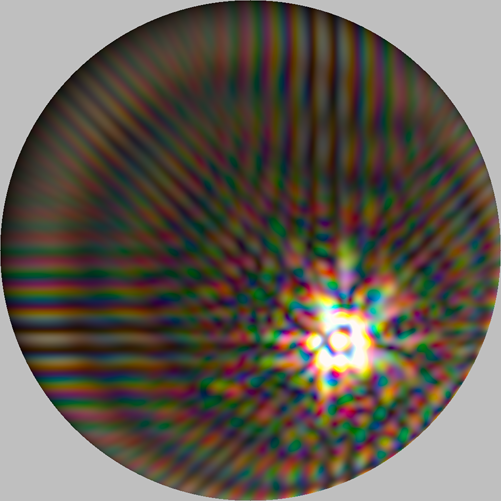} \\
    
    \rotatebox{90}{Predicted}&
    \includegraphics[width=0.14\linewidth]{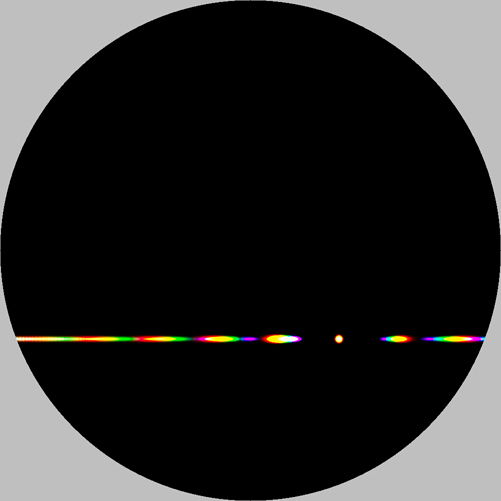} &
    \includegraphics[width=0.14\linewidth]{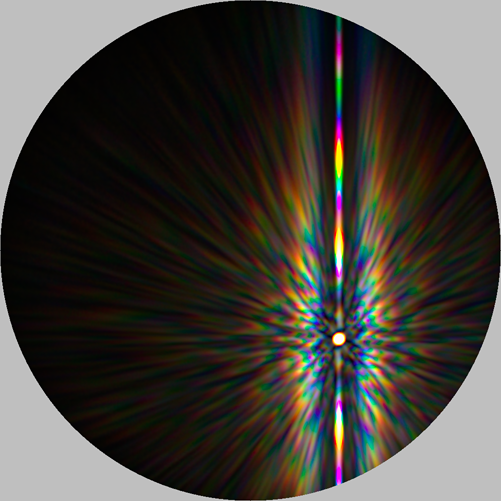} &
    \includegraphics[width=0.14\linewidth]{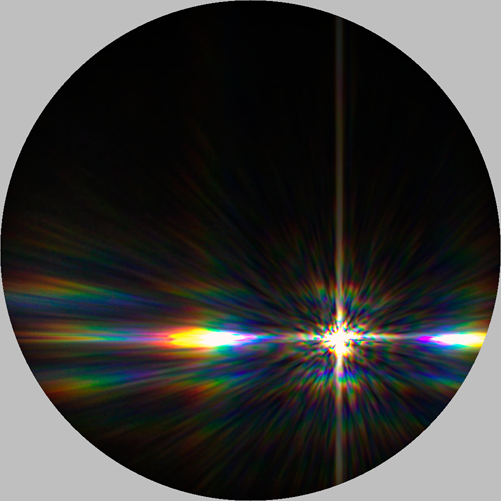} &
    \includegraphics[width=0.14\linewidth]{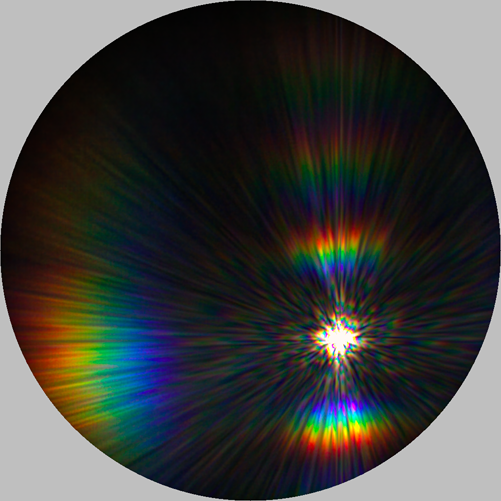} &
    \includegraphics[width=0.14\linewidth]{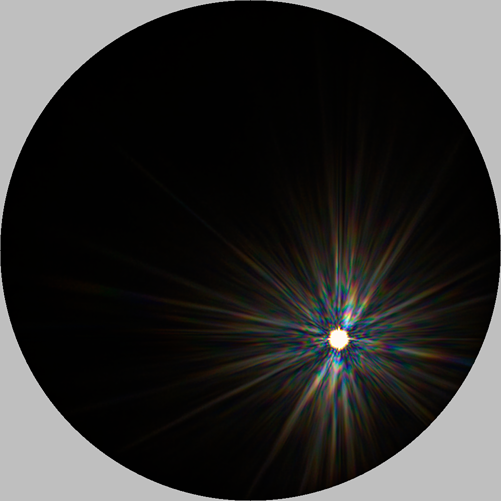} &
    \includegraphics[width=0.14\linewidth]{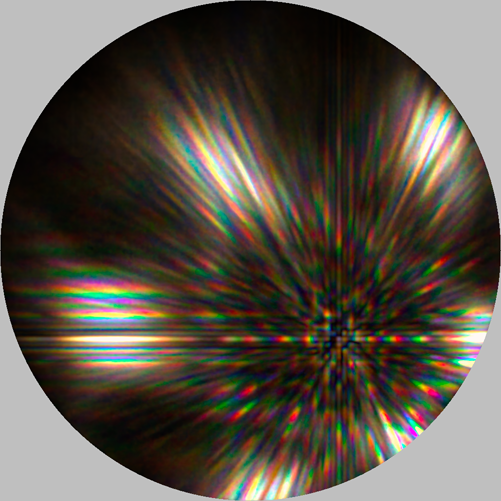} &
    \includegraphics[width=0.14\linewidth]{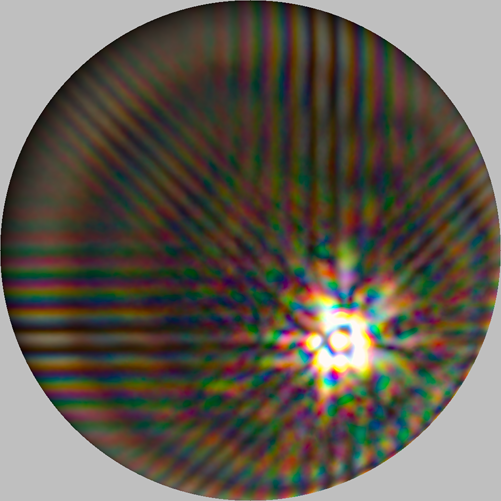} \\
    
    & Blazed Grating & Synthetic CD & Cornsnake & Sunbeam Snake & Ballpen Tip Scratch & Corner Cubes & Spherical Pits\\
    \end{tabular}

    \caption{BRDF slices for (\(\theta\), \(\phi\)) : \((30, 135)\) at the exposure \(\times 5000\). The predicted slices closely match the actual BRDF slices even at this high exposure level, demonstrating the robustness and accuracy of the proposed network.}
    \label{fig:slice30135}
\end{figure*}

\begin{figure*}[t!]
\small

    \centering
    \begin{tabular}{@{}c@{\hspace{1mm}}c@{}c@{}c@{}c@{}c@{}c@{}c}
    
    \rotatebox{90}{Actual}&
    \includegraphics[width=0.14\linewidth]{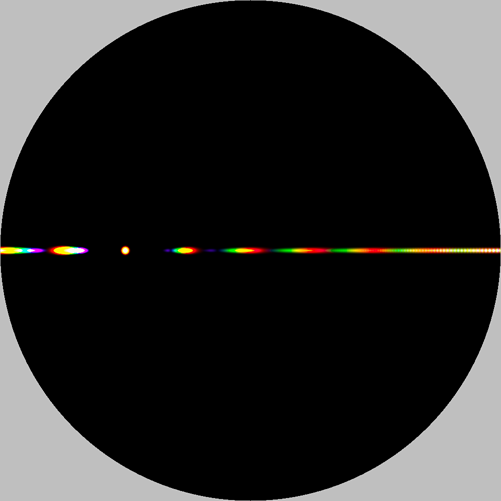} &
    \includegraphics[width=0.14\linewidth]{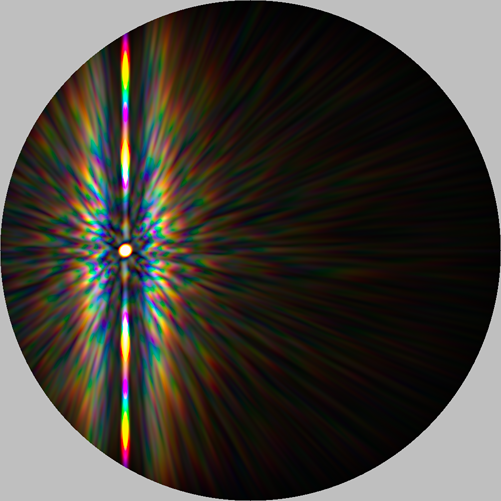} &
    \includegraphics[width=0.14\linewidth]{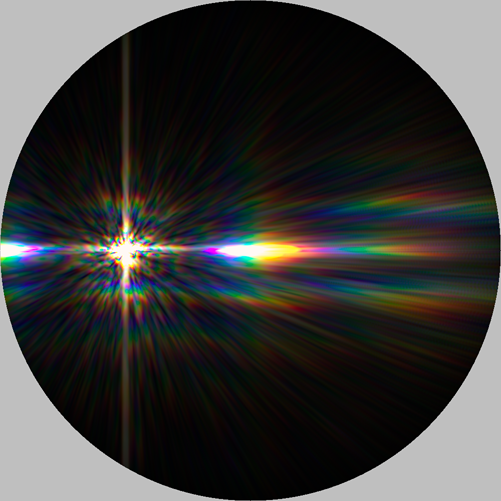} &
    \includegraphics[width=0.14\linewidth]{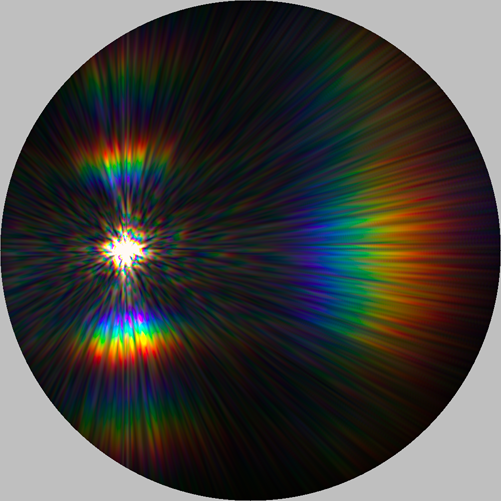} &
    \includegraphics[width=0.14\linewidth]{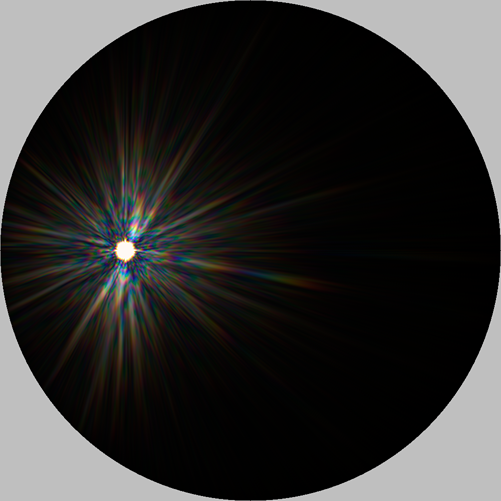} &
    \includegraphics[width=0.14\linewidth]{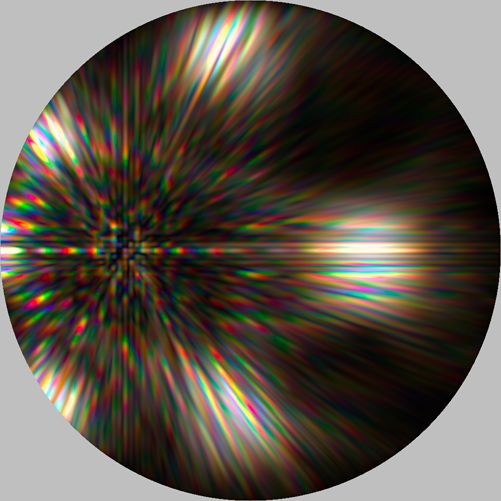} &
    \includegraphics[width=0.14\linewidth]{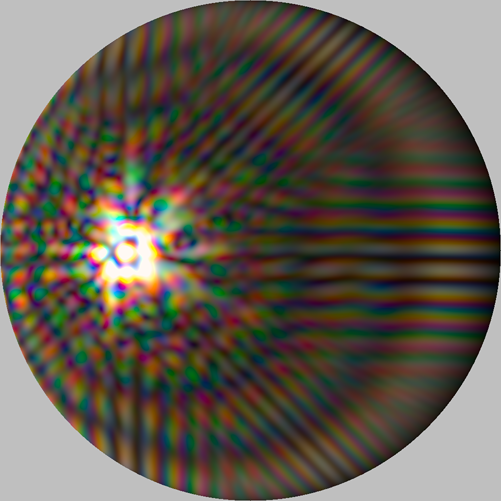} \\
    
    \rotatebox{90}{Predicted}&
    \includegraphics[width=0.14\linewidth]{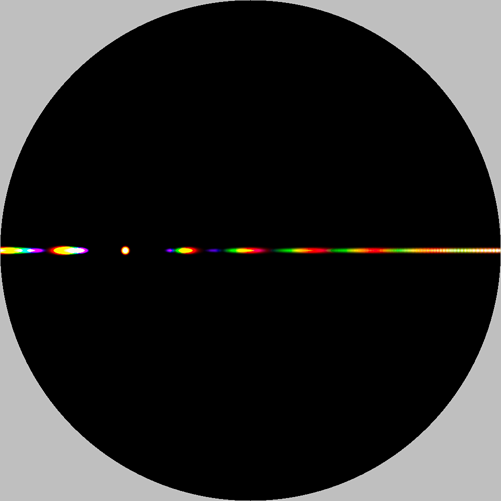} &
    \includegraphics[width=0.14\linewidth]{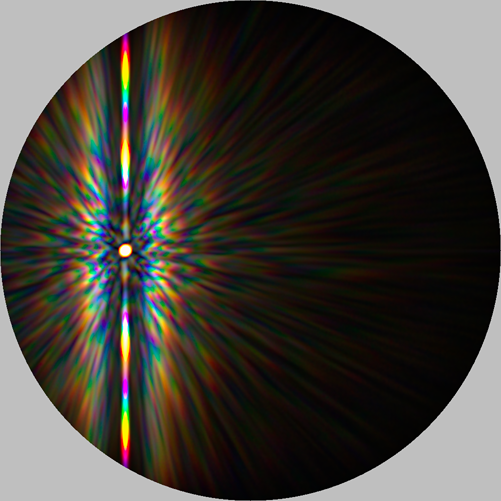} &
    \includegraphics[width=0.14\linewidth]{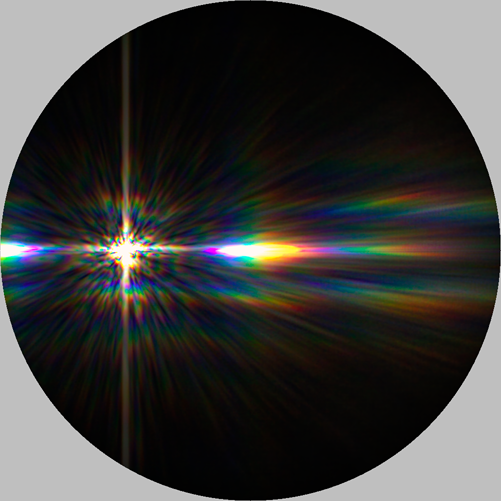} &
    \includegraphics[width=0.14\linewidth]{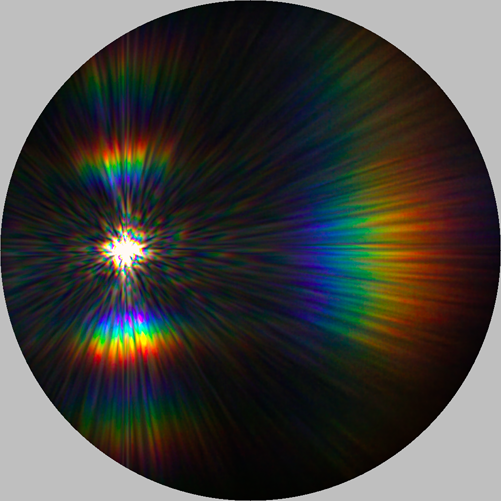} &
    \includegraphics[width=0.14\linewidth]{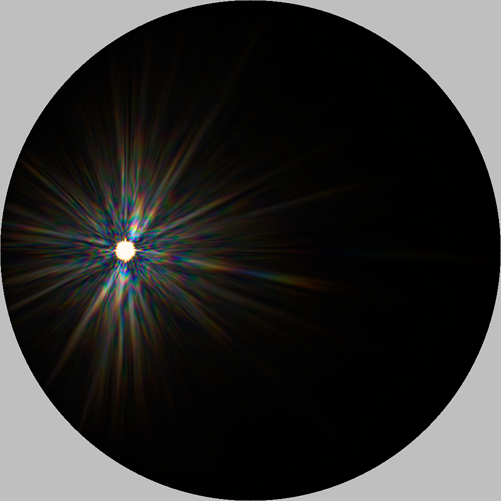} &
    \includegraphics[width=0.14\linewidth]{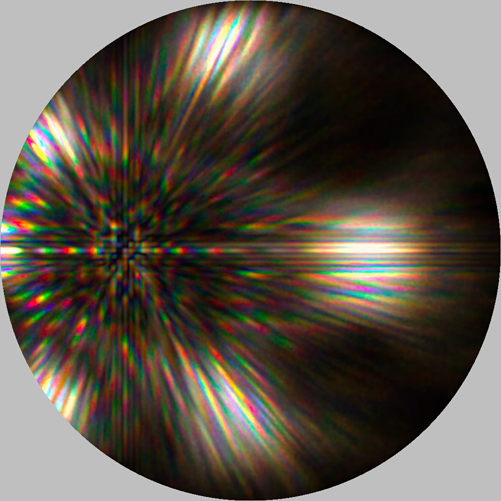} &
    \includegraphics[width=0.14\linewidth]{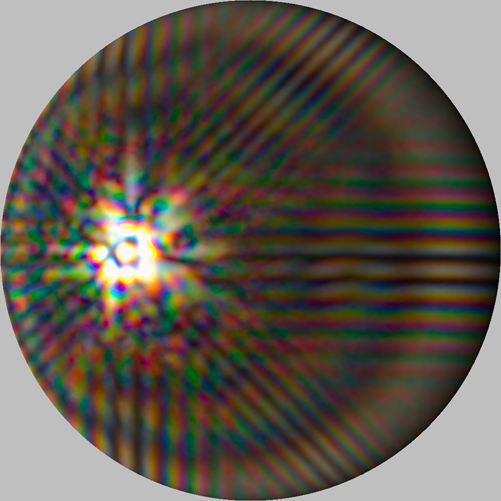} \\
    
    & Blazed Grating & Synthetic CD & Cornsnake & Sunbeam Snake & Ballpen Tip Scratch & Corner Cubes & Spherical Pits\\
    \end{tabular}

    \caption{BRDF slices for (\(\theta\), \(\phi\)) : \((30, 0)\) at the exposure \(\times 5000\). The predicted slices closely match the actual BRDF slices even at this high exposure level, demonstrating the robustness and accuracy of the proposed network.}
    \label{fig:slice300}
\end{figure*}

\begin{figure*}[t!]
\small

    \centering
    \begin{tabular}{@{}c@{\hspace{1mm}}c@{}c@{}c@{}c@{}c@{}c@{}c}
    
    \rotatebox{90}{Actual}&
    \includegraphics[width=0.14\linewidth]{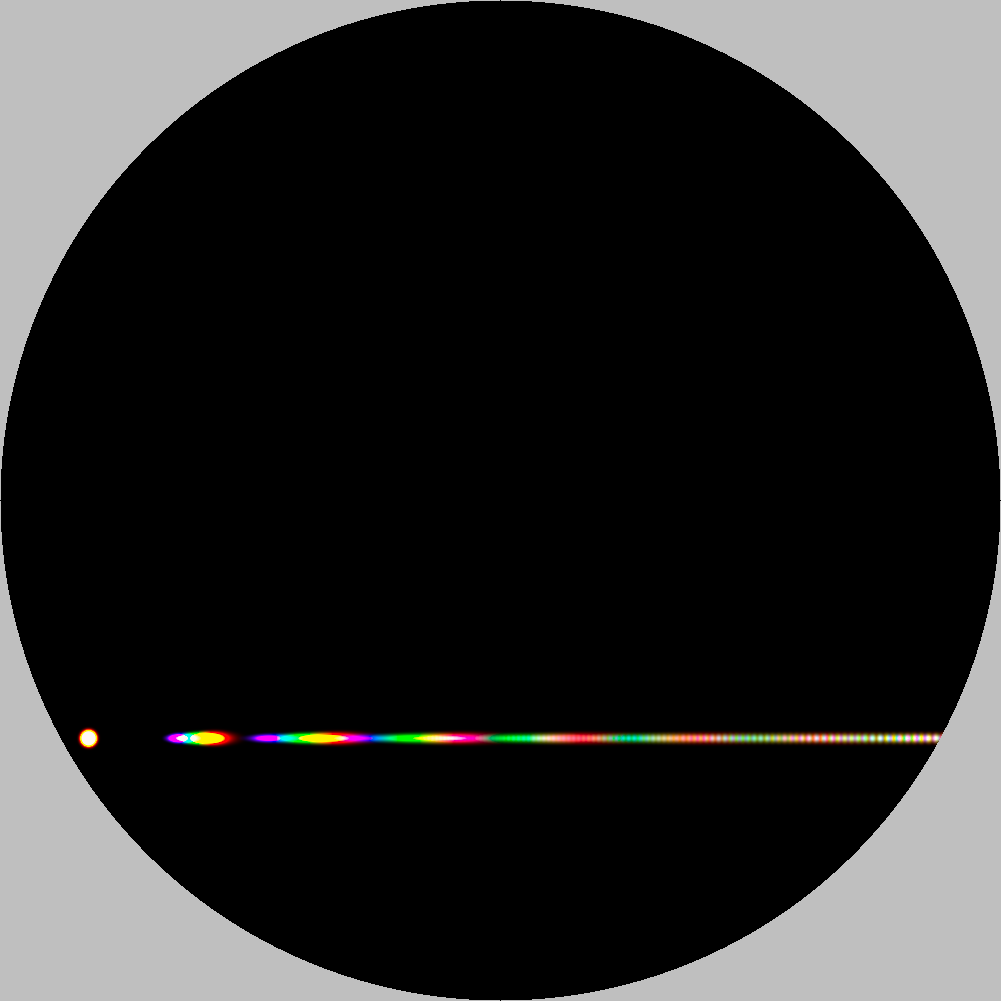} &
    \includegraphics[width=0.14\linewidth]{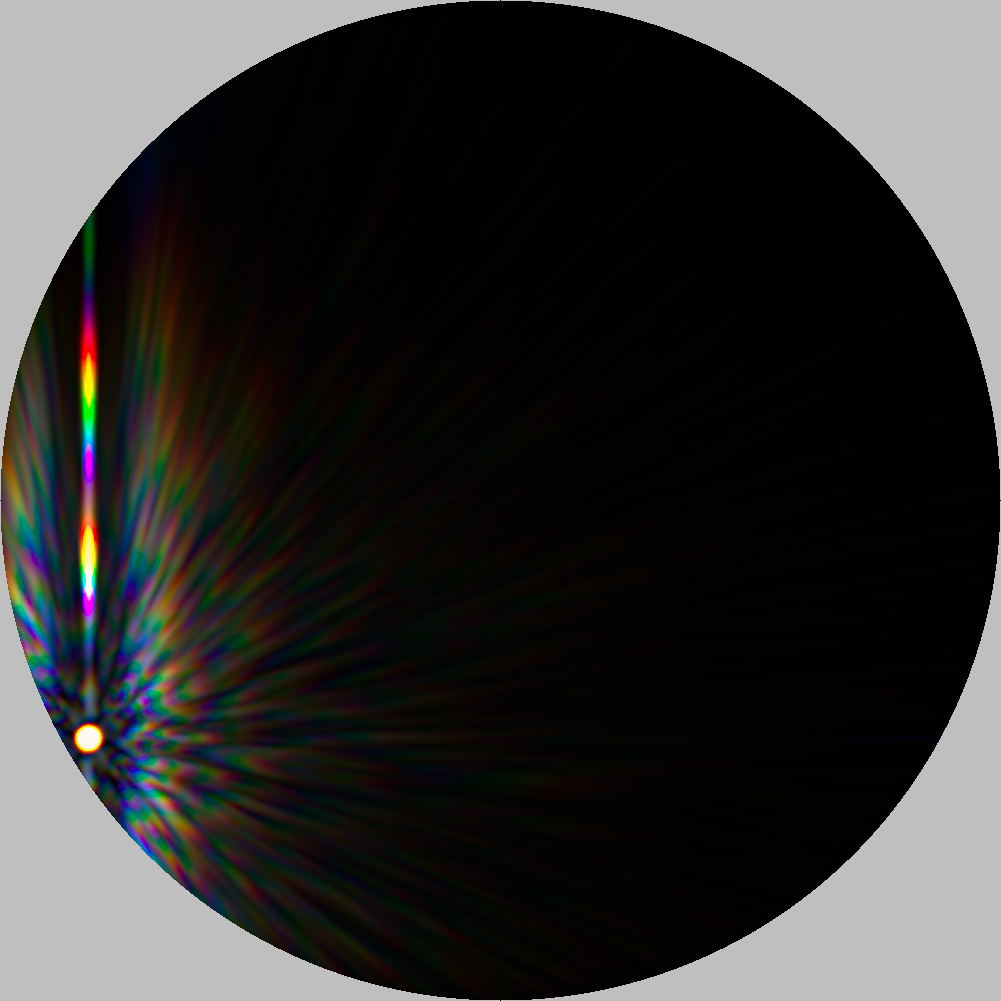} &
    \includegraphics[width=0.14\linewidth]{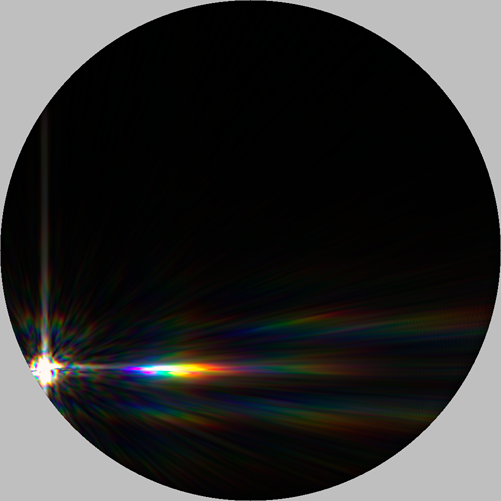} &
    \includegraphics[width=0.14\linewidth]{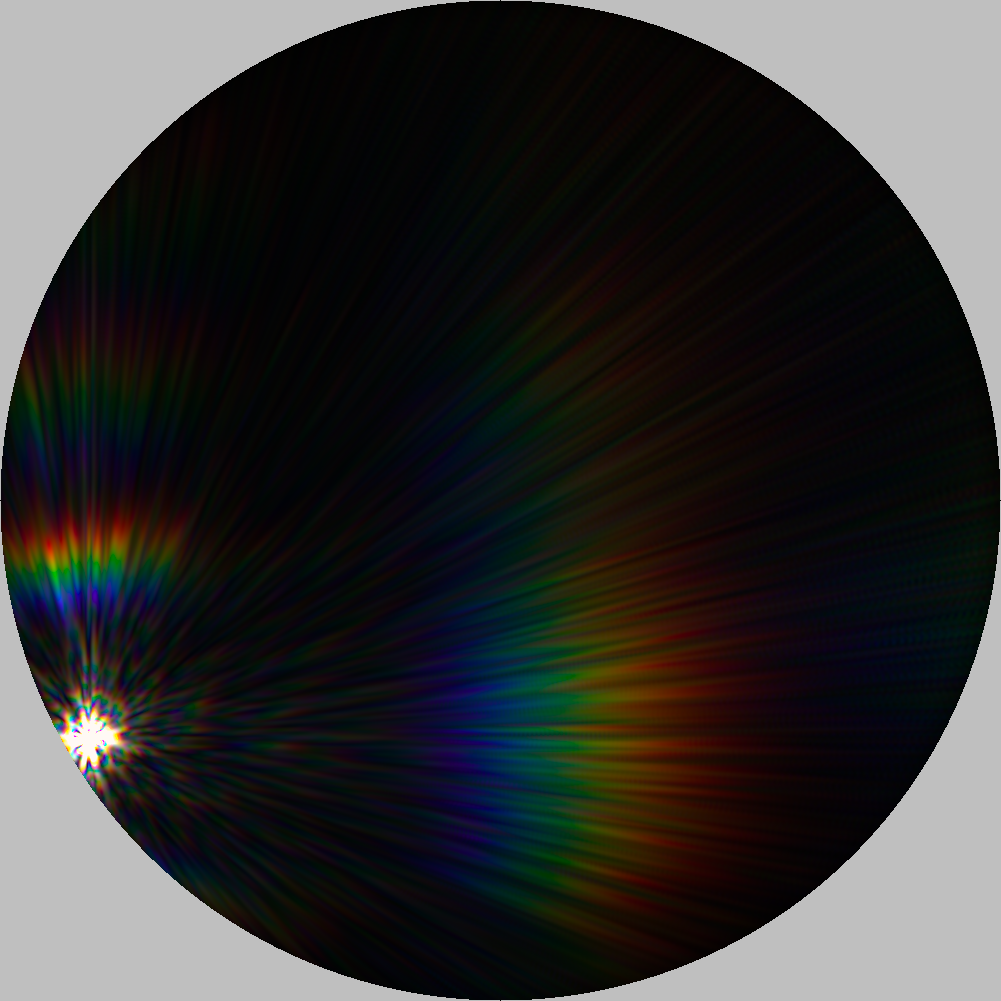} &
    \includegraphics[width=0.14\linewidth]{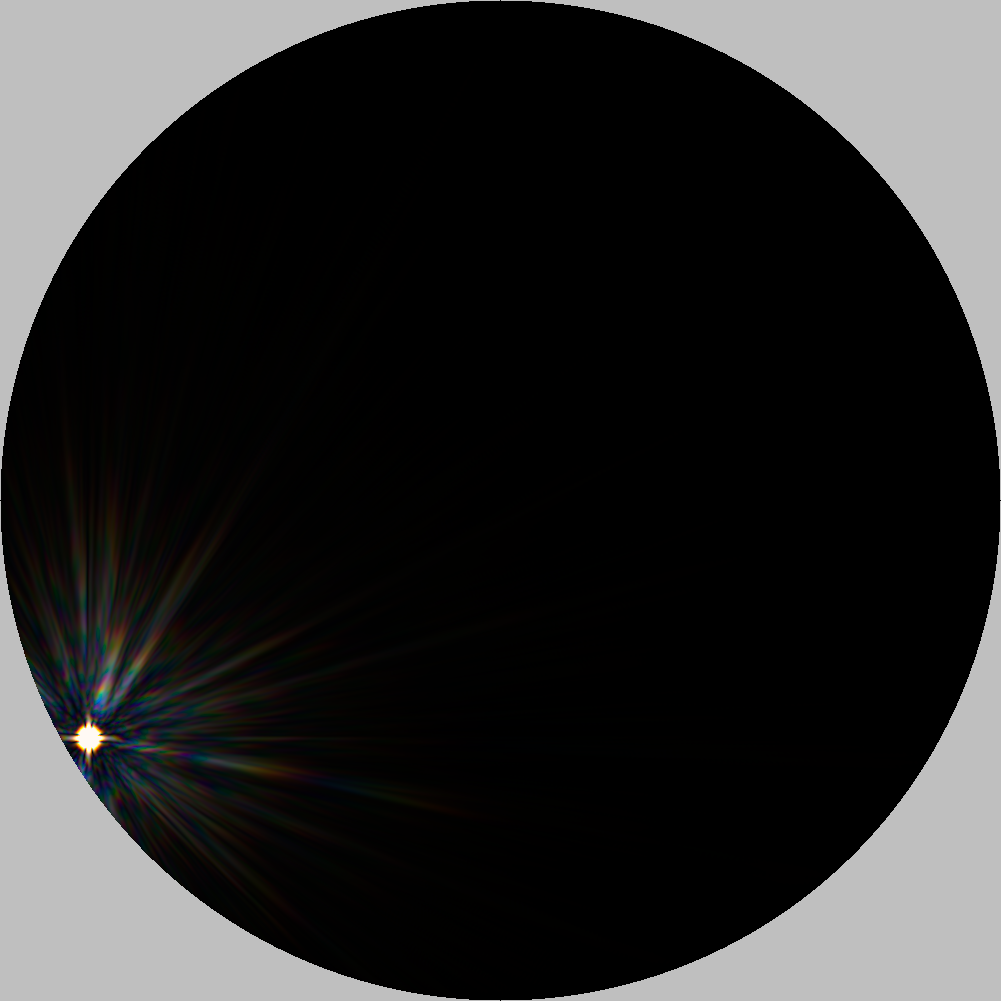} &
    \includegraphics[width=0.14\linewidth]{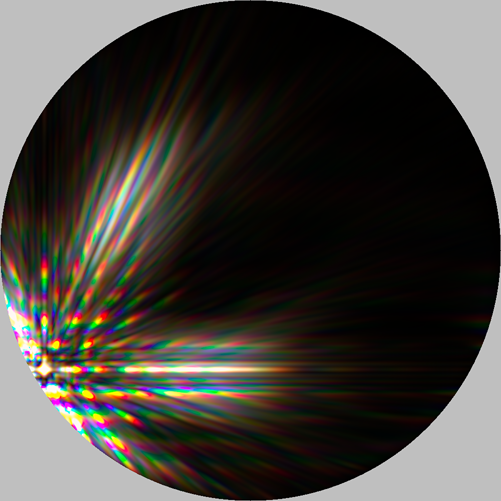} &
    \includegraphics[width=0.14\linewidth]{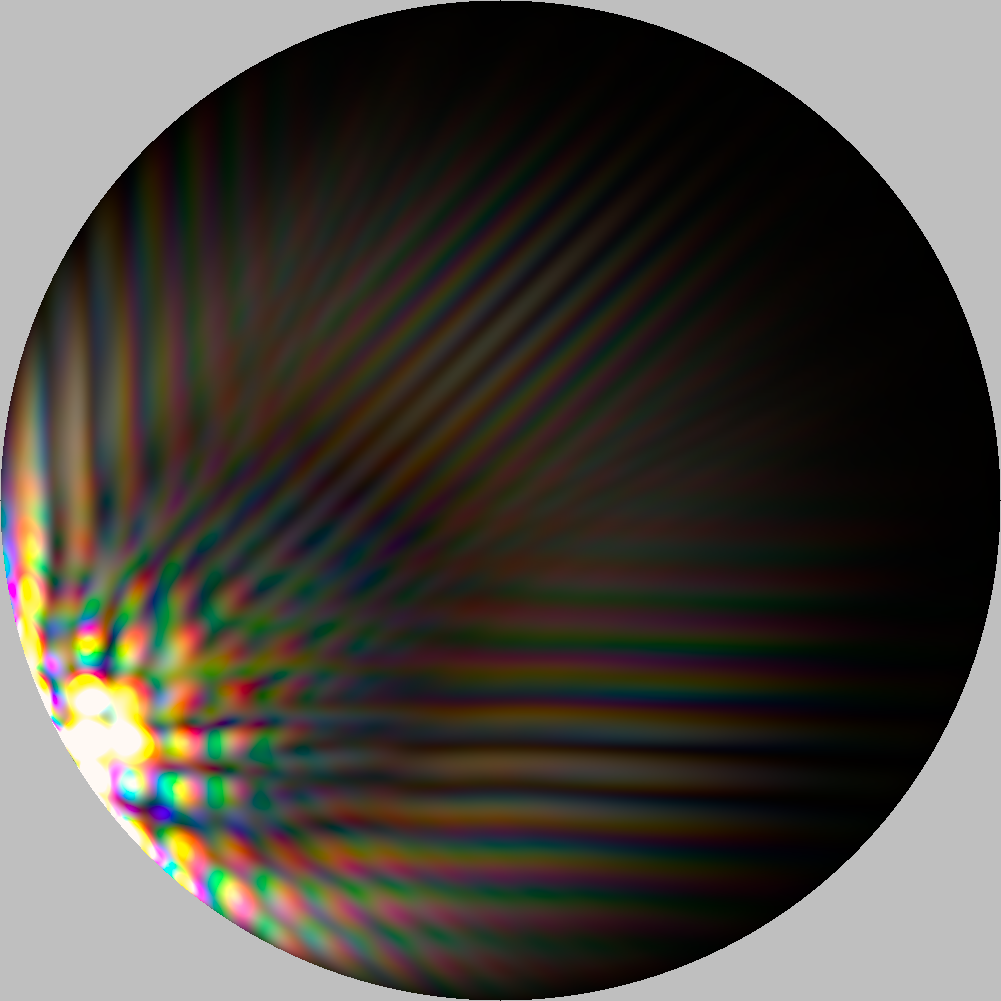} \\
    
    \rotatebox{90}{Predicted}&
    \includegraphics[width=0.14\linewidth]{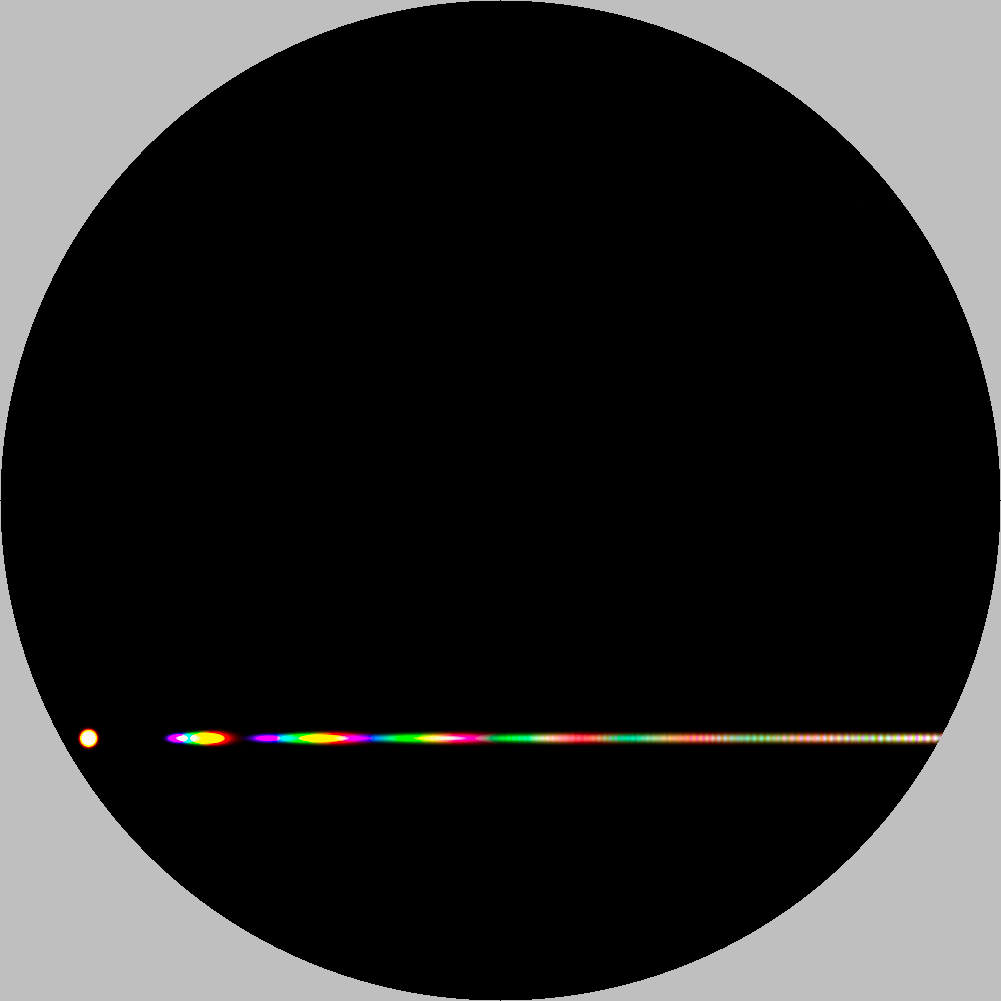} &
    \includegraphics[width=0.14\linewidth]{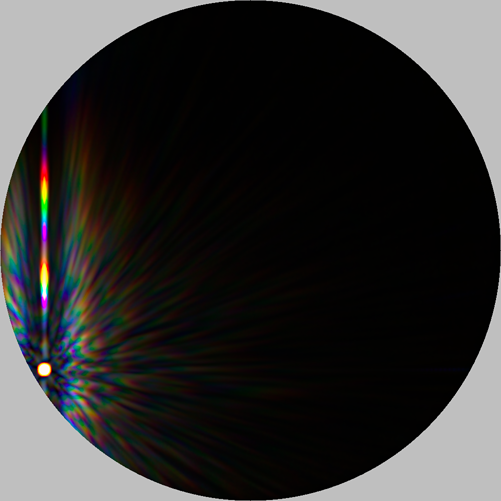} &
    \includegraphics[width=0.14\linewidth]{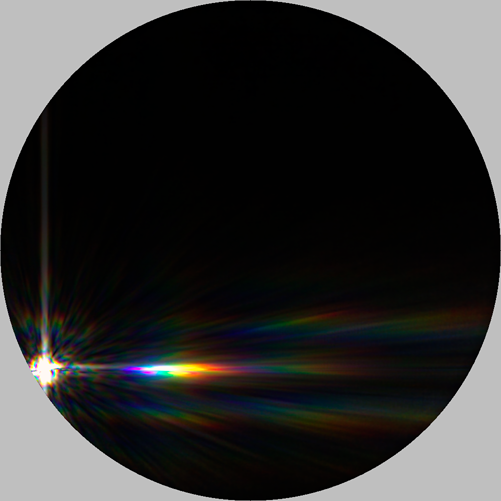} &
    \includegraphics[width=0.14\linewidth]{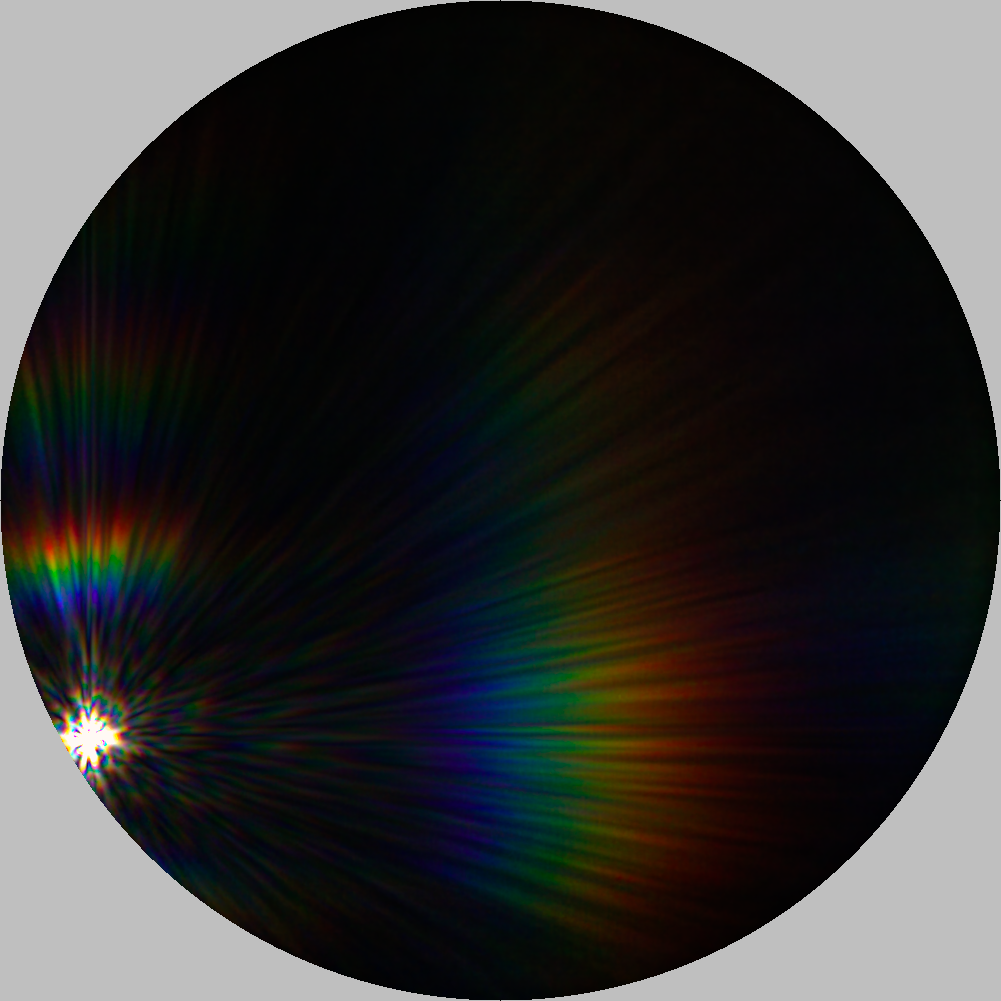} &
    \includegraphics[width=0.14\linewidth]{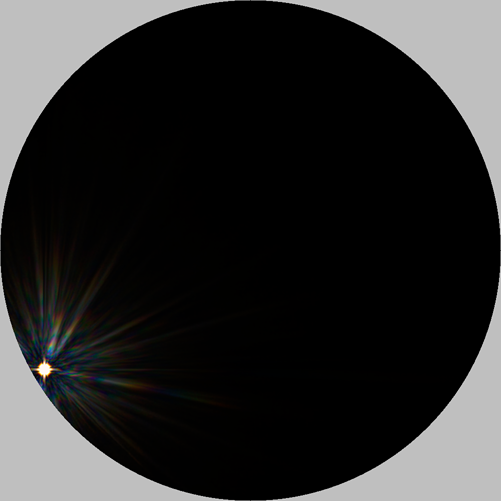} &
    \includegraphics[width=0.14\linewidth]{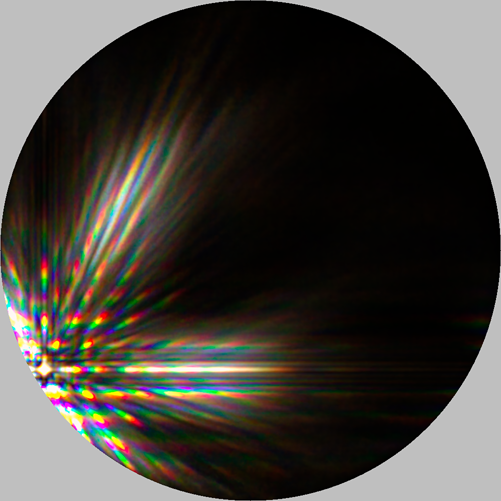} &
    \includegraphics[width=0.14\linewidth]{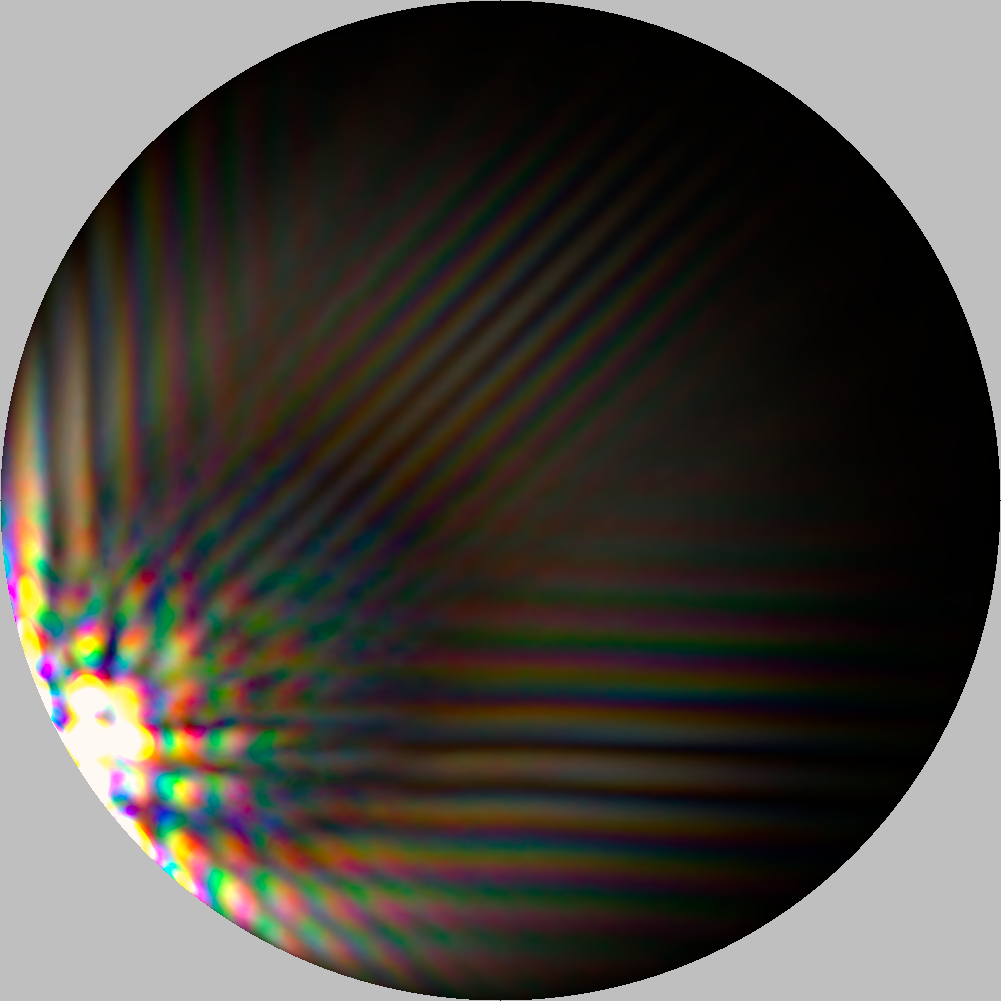} \\
    
    & Blazed Grating & Synthetic CD & Cornsnake & Sunbeam Snake & Ballpen Tip Scratch & Corner Cubes & Spherical Pits\\
    \end{tabular}

    \caption{BRDF slices for (\(\theta\), \(\phi\)) : \((72, 30)\) at the exposure \(\times 5000\). The predicted slices closely match the actual BRDF slices even at this high exposure level, demonstrating the robustness and accuracy of the proposed network.}
    \label{fig:slice7230}
\end{figure*}

\begin{figure*}[ht!]
\small
    \centering
    \begin{tabular}{c@{\hspace{1mm}}c@{}c@{}c@{}c@{}c@{}c@{}c}
    
    \rotatebox{90}{Actual}&
    \includegraphics[width=0.14\linewidth]{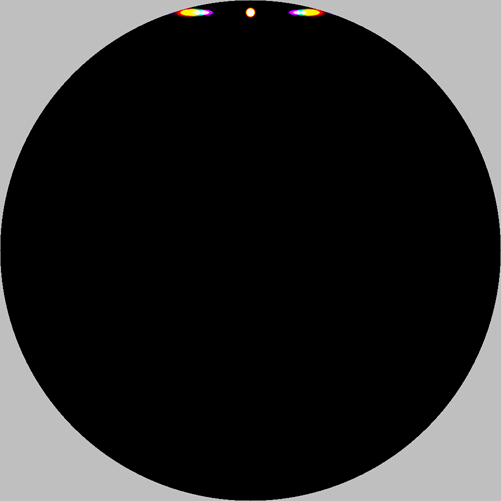} &
    \includegraphics[width=0.14\linewidth]{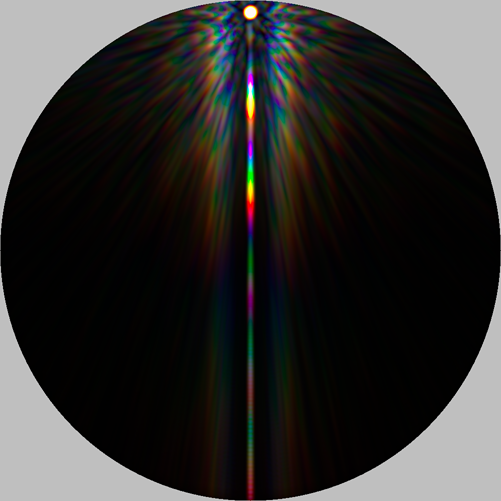} &
    \includegraphics[width=0.14\linewidth]{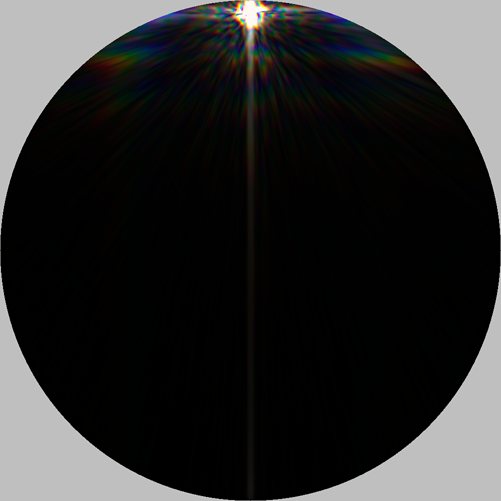} &
    \includegraphics[width=0.14\linewidth]{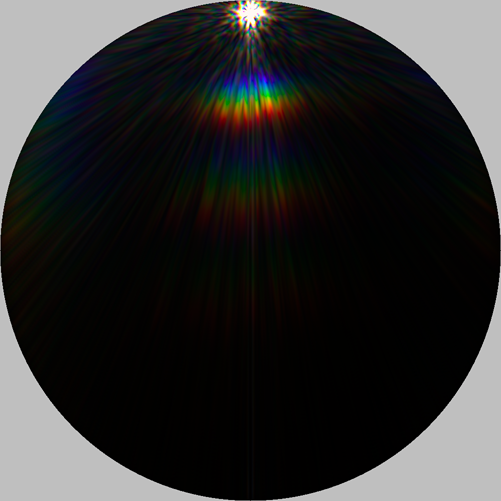} &
    \includegraphics[width=0.14\linewidth]{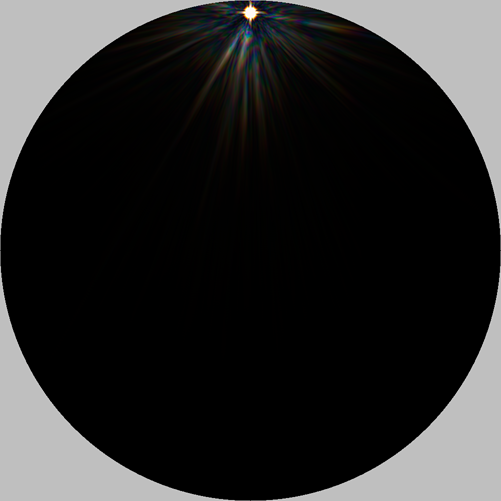} &
    \includegraphics[width=0.14\linewidth]{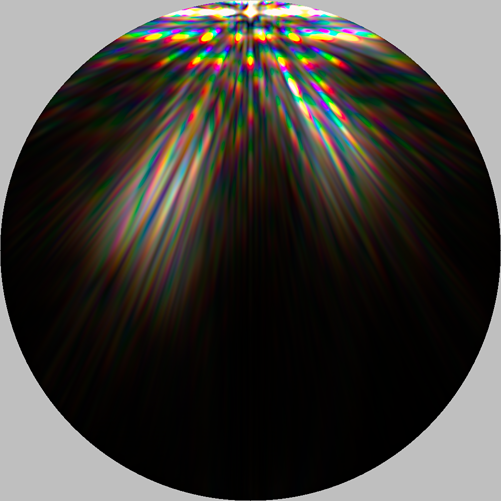} &
    \includegraphics[width=0.14\linewidth]{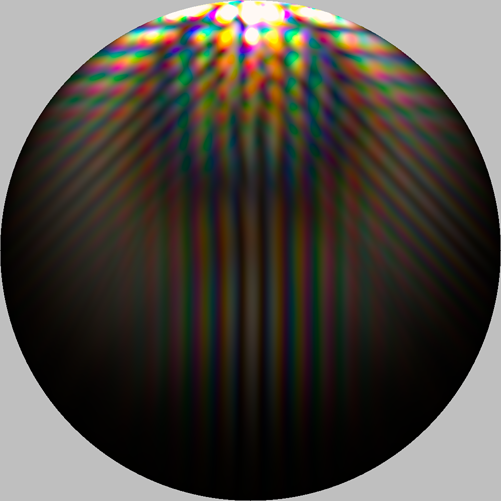} \\
    
    \rotatebox{90}{Predicted}&
    \includegraphics[width=0.14\linewidth]{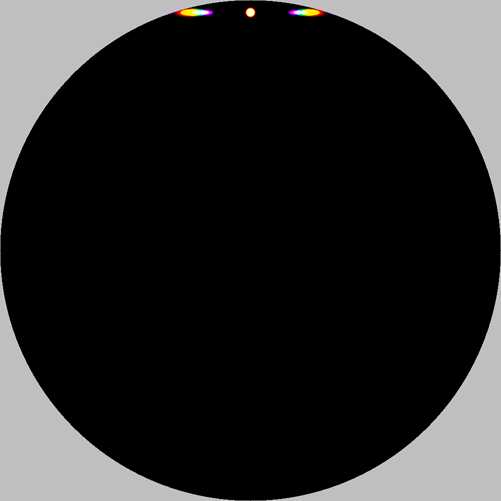} &
    \includegraphics[width=0.14\linewidth]{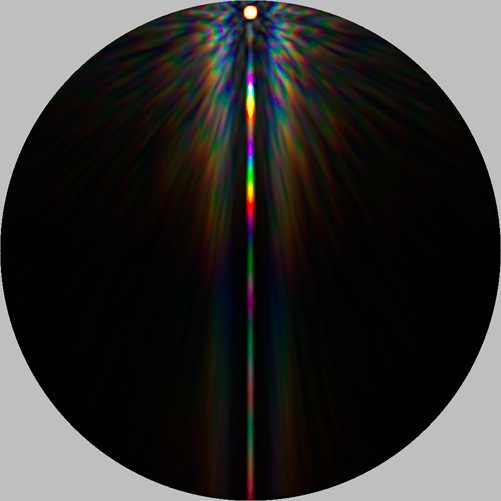} &
    \includegraphics[width=0.14\linewidth]{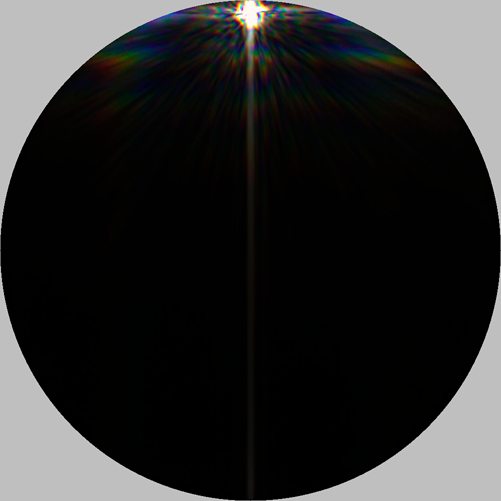} &
    \includegraphics[width=0.14\linewidth]{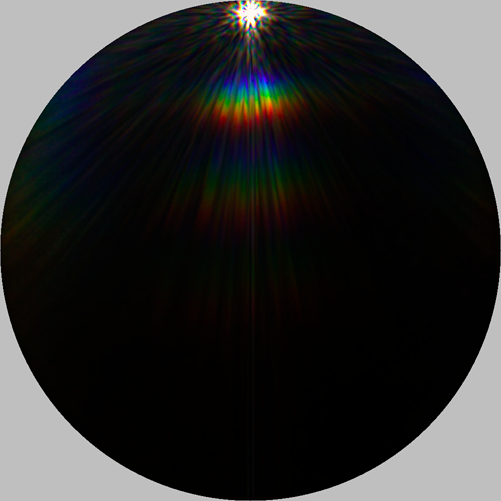} &
    \includegraphics[width=0.14\linewidth]{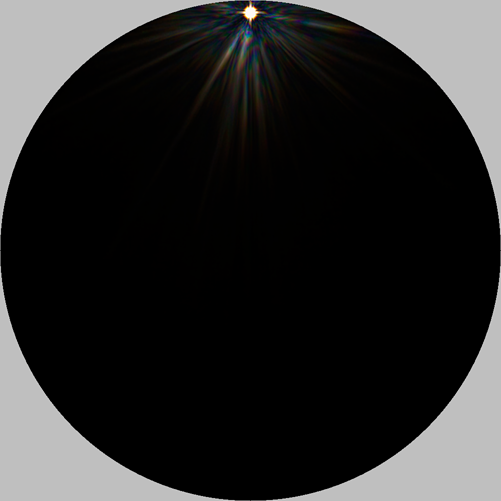} &
    \includegraphics[width=0.14\linewidth]{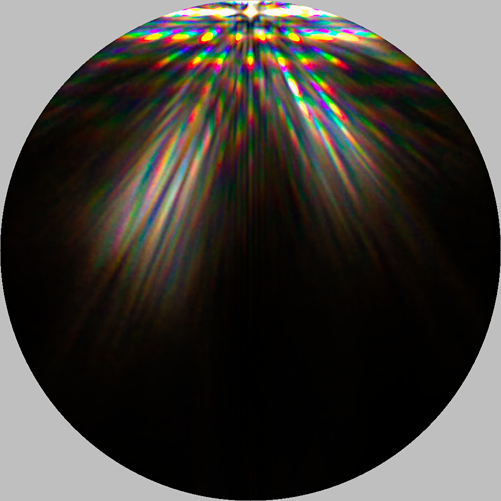} &
    \includegraphics[width=0.14\linewidth]{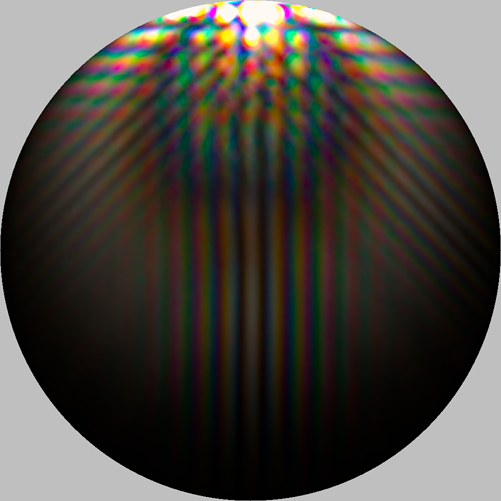} \\
    
    & Blazed Grating & Synthetic CD & Cornsnake & Sunbeam Snake & Ballpen Tip Scratch & Corner Cubes & Spherical Pits\\
    \end{tabular}

    \caption{BRDF slices for (\(\theta\), \(\phi\)) : \((72, 270)\) at the exposure \(\times 5000\).}
    \label{fig:slice72270}
\end{figure*}

Figures~\ref{fig:slice00_2000}, ~\ref{fig:slice30135}, ~\ref{fig:slice300} ~and ~\ref{fig:slice7230} show the results for our method for all the datasets at different (\(\theta\), \(\phi\)) at exposure $\times5000$. Visually our renderings look compellingly similar to the ground truths. Figures~\ref{fig:slice00_2000} depicts and also tabulates the \FLIP errors. The mean \FLIP error was generally found to lie in the range of $0.002$\textendash$0.087$ at $2000$RU. 

\begin{figure}[htb]
\includegraphics[width=0.99\linewidth]{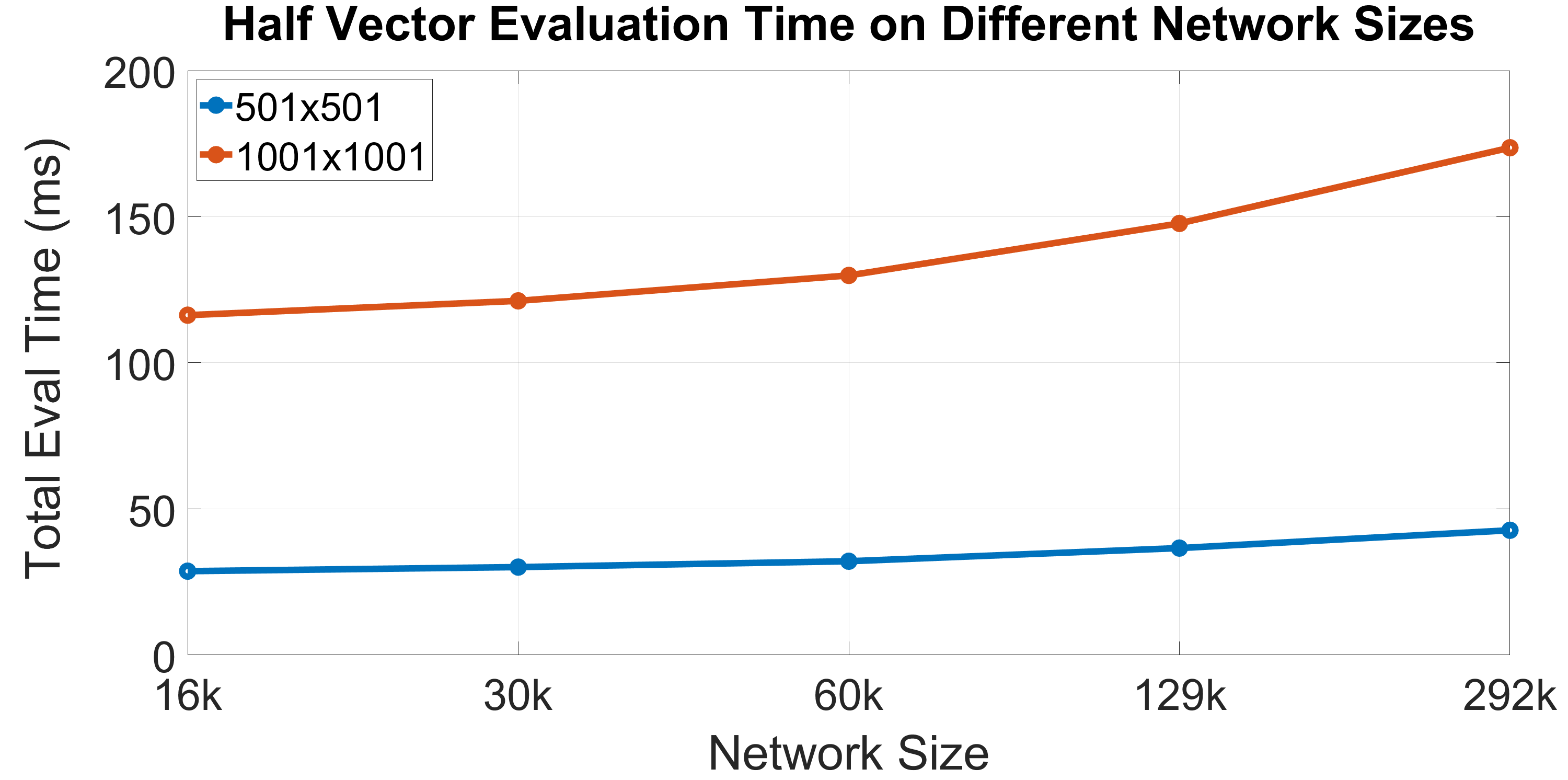}
\captionof{figure}{Performance comparison across different networks for a half-vector slice of size $501 \times 501$ and $1001 \times 1001$. Timing was recorded on a single NVIDIA RTX 3090 GPU. }
\label{fig:performance}
\end{figure}

\paragraph*{Performance Evaluation} We benchmarked our deferred shading algorithm implemented in Python to interoperate with the Mitsuba3 software~\cite{mitsuba3} that generates input AOV (arbitrary output variable) maps that drive our shader. These AOVs include position vectors and their gradients, surface normals, texture coordinates and their gradients, all projected into the screen space. We timed our MLP evaluator on a platform with one NVIDIA 3090 GPU. Figure~\ref{fig:performance} plots the performance graphs for two screen sizes for different network sizes. In general, our method takes $120ms$ to $175ms$ to render one frame of size $1001\times1001$. It can thus support rendering at $5-8$ frames-per-second. While these rates are not suitable for gaming or interactive applications, considering the high-quality output, our method can practically emulate the results of the recent full-wave forward simulator by~\citet{yu2023}. For the context, their forward method takes from $4.9$ to $17.3$ minutes for each such rendering using a platform with four NVIDIA $3090$ GPUs running simultaneously.    

\paragraph*{Memory Footprint}
Figure~\ref{fig:PsnrFlip}(b) shows that for the nanostructural complexity of groups \textit{basic} and \textit{real-world scans}, which also include natural gratings, high-quality reconstructions are achievable with a network of size $60K$ parameters stored as $4$ byte single precision floating numbers. The \FLIP errors in the range less than $0.04$ indicate good subjective matching. For \textit{complex designs} around $130K$ parameters generate such high subjective quality. Consider the interactive method from by ~\citet{dhillon2016} which requires a memory footprint of around $12$ million bytes in general with $4$ lookup tables of size $501\times501$ with $3$ RGB channels and each lookup entry of $4$ single-precision floating numbers. In comparison, our method produces footprint reduction of around $50\times$ for the \textit{basic} and \textit{real-world scans} and around $23\times$. Consider a direct comparison to the image compression method by \citet{catania2023nif} which requires $1$ to $1.5$ bits per pixel of $24$-bits. For $11$ million data samples of $3$ channels and $4$ bytes per channel, their compression should require a network of $1.375$ million parameters. In comparison, our method results in $10\times$ to $23\times$ further reductions. 

\paragraph*{Surface Renderings} Figure~\ref{fig:teaser} shows structural colors produced by our shader when it interoperates with the Mitsuba3 software \cite{mitsuba3} in a deferred rendering approach. For comparison, we also simulate a regular sinusoidal grating with $600$ lines-per-mm. We produce surface renderings for the most complex nanostructures (from ~\citet{yu2023}) at high efficiency and no visual artifacts.   

We also performed several ablation studies to empirically determine the comprehensive behavior of our method under different configurations. These studies mainly use PSNR and SSIM metrics and are detailed in \textit{Appendix B} in the \textit{supplemental material}.

\section{CONCLUSION}
\label{sec:Conclusion}

In this paper, we present a novel method for implicit neural representation of diffractive reflectances. Using the ground truth generated by a high-quality, Fourier optics-based reference forward modeler for Fraunhofer diffraction, our method is shown to reconstruct such reflectances with high subjective as well as objective quality. The PSNR for most of our reconstructions under the higher range of relative exposure ($2000$\textendash$5000$RU) is observed to be generally in the range of $45$\textendash$60$dB. The subjective \FLIP error measure ranges from $0.01$ to $0.09$ (lower being better) even for the most challenging nano-structural patches. These high-quality reconstructions are the consequence of the combinatorial influences of: (a) our proposed domain and range transformation schemes that are fine tuned to diffractive reflectances, (b) use of a funneling mechanism in the network design to smoothly annihilate learning losses, and (c) adaptations of the Fourier frequency based input expansion approach to explicitly infuse diagonal frequencies in the $u-v$ space for the half vector. For the first time, to the best of the author/s knowledge, diffractive surfaces of very high complexities as introduced by \citet{yu2023} are shown to be rendered accurately as well as efficiently. Our proposed method has strong potential to be incorporated into physically based global illumination pipelines for practical emulation of nuanced structural coloration due to diffraction.     

\section*{ACKNOWLEDGMENTS}
We sincerely thank Prof. Milinkovitch and colleagues from the University of Geneva for providing us with the atomic force microscopic (AFM) scans for the snake skins. We use these scans to represent gratings with natural variations in our experiments. We also thank ~\citet{yu2023} for sharing their complex nano-architectures and NanoSurf Corporation~\cite{nanoSurf} for sharing their AFM scans of a ball-pen tip, publicly. These datasets have been instrumental in evaluating our proposed method. 

This research used in part resources on the Palmetto Cluster at Clemson University under National Science Foundation awards MRI 1228312, II NEW 1405767, MRI 1725573, and MRI 2018069. The views expressed in this article do not necessarily represent the views of NSF or the United States government. 

\bibliography{gradnet}
\vfill{}
\clearpage
\pagebreak

\section*{Appendix A: Taylor Series Expansion for Approximating Diffraction BRDFs}
Surface nanostructures can be represented with a discrete 2D height-field function $h(x,y)$, representing nano elevations along the $z$-axis (surface normal). Fourier optical formulation for its spectral bidirectional reflectance distribution function (BRDF) is given as~\cite{dhillon2014}:
\begin{align}
    \mathit{f}(\lambda, \omega_i,\omega_o) = A(\omega_i,\omega_o)\left| \mathcal{F}\{p(x,y)\}\left(\dfrac{u}{\lambda},\dfrac{v}{\lambda}\right)\right|^2 ,
    \label{eqn:DTFT_Again}
\end{align}
where, $\lambda$ is the light wavelength, $\omega_i$ is the normalized incident light direction and $\omega_o$ is the normalized view direction. The unnormalized half vector $\omega_h = \omega_i + \omega_o$ is related to Stam's key vector $(u,v,w) = -\omega_h$~\cite{stam1999}. 
$\mathcal{F}$ represents Discrete-Time Fourier Transform (DTFT) of the phasor function $p(x,y) = e^{iwkh(x,y)}$, where $i$ is the imaginary identity, $k=2\pi/\lambda$ in the wave number, and  $w$ is the third element of the key vector. Lastly, $A(\omega_i,\omega_o)$ is a normalized, net attenuation term that incorporates the Fresnel term, geometric attenuation and the masking and shadowing effects. More specifically,
\begin{align}
    A(\omega_i,\omega_o) 
    &=R^2(\omega_i,\omega_o) G(\omega_i,\omega_o) / R_0^2 w^2, &\text{for } |w| > w_\epsilon, \nonumber\\
    &= 0 &\text{otherwise.}
\end{align}
Here, $R$ is the Fresnel term, and $R_0$ is the reference Fresnel term for normal incidence under normal viewing, i.e. $R(\omega_i = 0,\omega_o = 0)$. $G$ is the standard geometric term expressed as \begin{equation} G(\omega_i,\omega_o) = \frac{(1+\omega_i \omega_o)^2}{cos\theta_i cos\theta_o},\end{equation} with $\theta$ as the polar angle for the respective direction vector. $w_\epsilon$ is a threshold at the scale of $10^{-6}$ and used to emulate a basic masking and shadowing function for extreme grazing angles. 

To simplify computations for practical operations, \citet{dhillon2014} show that a physically-based Gaussian coherence windowing of the discrete phasor function $p(x,y)$ under Taylor series expansion yields to Discrete Fourier Transformations (DFTs). This simplification for $\mathcal{F}\{p(x,y)\}$ in Equation~\ref{eqn:DTFT_Again} is formulated as 
\begin{align}
    \mathcal{F}\{p\}(\dfrac{u}{\lambda},\dfrac{v}{\lambda}) = 
    \sum_{n=0}^{N}{\dfrac{(iwk)^n}{n!}} 
    (\mathscr{F}\{h^n\}\ast\mathscr{G})\left(\dfrac{u}{\lambda},\dfrac{v}{\lambda}\right).
    \label{eqn:Taylor}
\end{align}
Here, $\mathscr{F}\{h^n(x,y)\}$ is the discrete Fourier transformation of the height-field $h(x,y)$ raised to power $n$, element-wise. $\mathscr{G}(u/\lambda,v/\lambda)$ is the frequency-space equivalent of a given physically-based spatial coherence area. This coherence area limits wave interference and is modeled as an isotropic Gaussian with given standard deviation $\sigma_s$ in microns. Using $\sigma_s$, we can derive $\mathscr{G}$ as 
\begin{align}
    \mathscr{G}(\xi,\psi) = \exp\left(- \dfrac{\xi^2 + \psi^2}{2 \sigma_f}\right), \quad\text{where } \sigma_f = \dfrac{1}{2\pi\sigma_s}. 
\end{align}
$\mathcal{G}$ spans only a small window in the frequency space and emulating Equation~\ref{eqn:DTFT_Again} by substituting Equation~\ref{eqn:Taylor} in it is computationally faster and much practical. Lastly, $N$ is kept arbitrarily large to ensure the convergence of the Taylor series. By default, we perform spectral sampling at $5$nm resolution and collapse resulting spectral response subject to $D65$ illumination under spectral integration. Resulting map for the $\mathcal{F}$-term in CIE-XYZ colorspace. For brevity, we still refer to it as the reflectance map. 

\section*{Appendix B: Ablation Studies}
\label{sec:ablations}
For the ablation studies, we focused on networks of smaller sizes since those configurations particularly challenge the compressive capabilities of our method and thus offer rich learning opportunities.

After numerous experiments, we empirically set the default challenging configuration to have fifteen frequencies corresponding to $u$ and $v$ dimensions and four relating to $w$ dimension. We use the input expansion scheme given by Equation (8) in the main paper.
In this following, we present the ablation studies on \textit{logDivisor}, \textit{range space transformation}, \textit{frequency count} and \textit{network size}. We base our analysis on PSNR and SSIM errors. 

\paragraph*{Range-Space Transformation}
We performed experiments with different data transformation techniques. The PSNR and SSIM for different range-space transformation is shown in the table~\ref{tab:ln_range_trans}. This data clearly shows that using range transformations with Equation. 6 from the main paper results in best quantitative results (highest PSNR and SSIM values). 

\begin{table}[b]
    \centering
\begin{tabular}{|c|c|c|c|c|}
\hline
\multirow{2}{*}{Transformation} & \multicolumn{2}{c|}{PSNR} & \multicolumn{2}{c|}{SSIM} \\
\cline{2-5}
& XYZ & lRGB & YBR & sRGB \\
\hline
$x$ & 31.781 & 29.810 & 0.853 & 0.420 \\
\hline
$log_2(1 + x)$ & 31.950 & 30.048 & 0.862 & 0.476 \\
\hline
 $1 + log_2 (x)/15$ & 30.437 & 30.335 & 0.717 & 0.284 \\
\hline
 $1 + log_2 (x)/20$ & 35.934 & 34.377 & 0.974 & 0.732 \\
\hline
 $(1 + log_2 (x)/24)^{2.5}$ & 36.090 & 35.923 & 0.981 & 0.927 \\
\hline
 $(1 + log_2 (x)/48)^{8}$ & \textbf{36.389} & \textbf{36.624} & \textbf{0.981} & \textbf{0.928} \\

\hline
\end{tabular}
\caption{PSNR and SSIM values for different range-space transformation}
\label{tab:ln_range_trans}
\end{table}

\begin{table}
\centering
\begin{tabular}{|c|c|c|c|c|c|}
\hline
\multicolumn{2}{|c|}{Parameters} & \multicolumn{2}{c|}{PSNR} & \multicolumn{2}{c|}{SSIM} \\
\cline{1-6}
$B_\text{max}$ & $n$ & XYZ & lRGB & YBR & sRGB \\
\hline
15 & 1 & 30.437 & 30.335 & 0.717 & 0.284 \\
15 & 2.5 & 30.089 & 29.615 & 0.692 & 0.254 \\
\hline
20 & 1 & 35.934 & 34.377 & 0.974 & 0.732 \\
20 & 2.5 & \textbf{36.418} & \textbf{36.840} & 0.976 & 0.703 \\
\hline
24 & 1 & 35.495 & 33.560 & 0.972 & 0.918 \\
24 & 2.5 & 36.090 & 35.923 & 0.981 & 0.927 \\
24 & 3.5 & 36.145 & 36.584 & 0.980 & 0.913 \\
\hline
30 & 2.5 & 36.006 & 35.286 & 0.979 & 0.927 \\
30 & 3.5 & 36.185 & 36.518 & 0.981 & 0.931 \\
\hline
45 & 2.5 & 35.906 & 35.099 & 0.975 & 0.920 \\
45 & 3.5 & 36.356 & 35.061 & 0.978 & 0.925 \\
\hline
48 & 8.0 & 36.389 & 36.624 & \textbf{0.981} & \textbf{0.928} \\
\hline
\end{tabular}
\caption{PSNR and SSIM values for different configurations}
\label{tab:ln-psnr-ssim}
\end{table}

\paragraph*{logDivisor}
The recorded PSNR and SSIM for different combinations of \textit{logDivisor} and power function are shown in Table~\ref{tab:ln-psnr-ssim}. While the PSNR is highest for $B_\text{max}=20$ with $n=2.5$, the SSIM which is a better predictor of the subjective quality is still the best for  $B_\text{max}=48$ with $n=8.0$. More importantly, higher the value of $B_\text{max}$, greater is the bit-precision and better reconstruction at higher exposures.




\paragraph*{Frequency Count}
From our experiments, we found that expanding input coordinates using sine and cosine functions helps the network generalize over high-frequency components only until the Nyquist sampling rates are reached (i.e. $19$ along $u$ and $v$ dimensions with $s= 1.4$ for the $u$-$v$ resolution of $1001\times1001$). In our ablation study, we compare PSNR and SSIM across various frequencies applied to the $u$, $v$, and $w$ coordinates. 

\paragraph*{Network Size}
We train the Synthetic CD dataset with an input size of 3 and a frequency of $(15, 15, 4)$ across various network architectures, varying the number of parameters from $1.5k$ to $245k$. From our experiments, we found that a network with hidden layers of size $(108, 66, 40, 24)$ achieves good performance. The PSNR and SSIM values for slice 7 with an exposure of $\times2640$ are presented in Table~\ref{tab:network-size}.
\vfill{}

\begin{table}[t]
\centering
\setlength{\tabcolsep}{3pt}
\begin{tabular}{|c|c|cc|cc|}
\hline
\multicolumn{2}{|c|}{Frequency} & \multicolumn{2}{c|}{PSNR} & \multicolumn{2}{c|}{SSIM} \\
\cline{1-6}
(U,V,W) & i/p size& XYZ & lRGB & YBR & sRGB \\
\hline
(0,0,0) & 3 & 34.064 & 31.862 & .957 & .872 \\
\hline
(5,5,5) & 33 & 34.815 & 32.680 & .965 & .889 \\
\hline
(10,10,5) & 53 & 35.726 & 35.027 & .976 & .911 \\
\hline
(15,15,3) & 69 & 36.502 & 36.076 & .980 & .927 \\
(15,15,4) & 71 & 36.090 & 35.923 & .981 & .927 \\
(15,15,5) & 73 & 36.093 & 36.086 & .982 & .929 \\
(15,15,15) & 93 & 35.896 & 36.077 & .982 & .928 \\
\hline
(20,20,5) & 93 & 36.094 & 36.332 & .981 & .928 \\
\hline
(30,30,5) & 143 & 36.396 & 36.425 & .982 & .928 \\
\hline
\end{tabular}
\caption{PSNR and SSIM values for different frequency configurations}
\label{tab:freq_metrics}
\end{table}
\begin{table}[H]
\centering
\setlength{\tabcolsep}{3pt}
\begin{tabular}{|c|c|cc|cc|}
\hline
\multicolumn{2}{|c|}{Hidden Layer} & \multicolumn{2}{c|}{PSNR} & \multicolumn{2}{c|}{SSIM} \\
\cline{1-6}
Node & Params & XYZ &  lRGB & YBR & sRGB \\
\hline
(10,10) & 863 & 31.140 & 27.137 & .938 & .828 \\
(21,21) & 2,040 & 33.425 & 31.014 & .950 & .850 \\
(10,21,21,10) & 1,666 & 33.488 & 31.719 & .951 & .852 \\
\hline
(108,66,40,24) & 18,709 & 36.090 & 35.923 & .981 & .927 \\
(120,48,32,24) & 16,883 & 35.822 & 35.879 & .979 & .924 \\
\hline
(280,100,48,16) & 53,943 & 36.371 & 37.226 & .990 & .958 \\
(360,150,70,32) & 93,011 & 36.374 & 37.431 & .994 & .969 \\
(450,220,85,32) & 153,256 & 36.358 & 37.422 & .995 & .976 \\
(680,260,65,30) & 245,058 & \textbf{36.435} & \textbf{37.612} & \textbf{.996} & \textbf{.979} \\
\hline
\end{tabular}
\caption{PSNR and SSIM values for different network configurations for slice $7$ at an exposure of $2640$}
\label{tab:network-size}
\end{table}

\end{document}